\documentclass[12pt]{article}
\usepackage{a4wide}
\usepackage{amssymb}
\usepackage{amsmath}
\usepackage{graphicx}

\begin{document}
{\renewcommand{\thefootnote}{\fnsymbol{footnote}}
\begin{center}
{\LARGE Effective potentials from semiclassical truncations}\\
\vspace{1.5em}
Bekir Bayta\c{s},\footnote{e-mail address: {\tt bub188@psu.edu}}
Martin Bojowald\footnote{e-mail address: {\tt bojowald@gravity.psu.edu}}
and
Sean Crowe\footnote{e-mail address: {\tt stc151@psu.edu}}
\\
\vspace{0.5em}
Department of Physics,
The Pennsylvania State
University,\\
104 Davey Lab, University Park, PA 16802, USA\\
\vspace{1.5em}
\end{center}
}

\setcounter{footnote}{0}

\begin{abstract}
  Canonical variables for the Poisson algebra of quantum moments are
  introduced here, expressing semiclassical quantum mechanics as a canonical
  dynamical system that extends the classical phase space. New realizations
  for up to fourth order in moments for a single classical degree of freedom
  and to second order for a pair of classical degrees of freedom are derived
  and applied to several model systems. It is shown that these new canonical
  variables facilitate the derivation of quantum-statistical quantities and
  effective potentials. Moreover, by formulating quantum dynamics in classical
  language, these methods result in new heuristic pictures, for instance of
  tunneling, that can guide further investigations.
\end{abstract}

\section{Introduction}

Semiclassical physics can often be described by classical equations of motion
amended by correction terms and possible new degrees of freedom. For instance,
Ehrenfest's theorem shows that the expectation values of position and momentum
in an evolving quantum state obey equations of motion which are identical with
the classical equations to zeroth order in $\hbar$ but, in general, have a
modified quantum force given by $-\langle \nabla V(\hat{x})\rangle$ not equal
to the classical force $-\nabla V(\langle\hat{x}\rangle)$ evaluated at
$\langle\hat{x}\rangle$. The difference depends on $\langle\hat{x}\rangle$,
but also on the variance $(\Delta x)^2$ and higher moments, which constitute
new, non-classical degrees of freedom.

A moment expansion can be used to derive quantum corrections
systematically. In this way, one can formulate quantum dynamics as
classical-type dynamics on an extended phase space, given by expectation
values and moments equipped with a Poisson bracket that follows from the
commutator of operators \cite{EffAc,Karpacz}. Moments, however, do not
directly form canonical variables on this Poisson manifold, which complicates
some of the usual procedures of canonical mechanics. Darboux' theorem
guarantees the existence of local canonical coordinates, but it is not always
easy to find them. Using a procedure we developed in \cite{Bosonize}, as well
as other new methods, we present here detailed derivations of canonical
variables for moments of up to fourth order for a single degree of freedom, as
well as to second order for a pair of degrees of freedom. The resulting
expressions can be used to make interesting observations about the behavior of
states, and they are crucial for the derivation of effective potentials. We
present several applications, including tunneling which is also discussed in
more detail in \cite{Ionization}.

\section{Canonical Effective Methods}

We use a quantum system of $N$ degrees of freedom with basic operators
$\hat{q}_j$ and $\hat{\pi}_k$, $1\leq j,k\leq N$ that are canonically
conjugate,
\begin{equation}
 [\hat{q}_j,\hat{\pi}_k] = i\hbar \delta_{jk}\,.
\end{equation}
In a semiclassical truncation \cite{EffAc,Karpacz}, the state space is
described by a finite-dimensional phase space with coordinates given by the
basic expectation values $q_j=\langle \hat{q}_j\rangle$ and
$\pi_k=\langle\hat{\pi}_k\rangle$ and, for positive integers $k_i$ and $l_i$
such that $\sum_{i=1}^N(k_i+l_i)\geq 2$, the moments
\begin{equation} \label{moments}
 \Delta\left(q_1^{k_1}\cdots q_N^{k_N}\pi_1^{l_1}\cdots \pi_N^{l_N}\right) =
 \langle (\hat{q}_{1} - q_{1})^{{k_1}} \cdots (\hat{q}_{N} - q_{N})^{{k_N}}
 (\hat{\pi}_{1} - \pi_{1})^{{l_1}} \cdots (\hat{\pi}_{N} - \pi_{N})^{{l_{N}}}
 \rangle_{\mathrm{Weyl}}  
\, , 
\end{equation}
where the product of operators is Weyl (totally symmetrically) ordered. 
The phase-space structure is defined by the Poisson bracket
\begin{equation} \label{Poisson}
\{ \langle \hat{A} \rangle, \langle \hat{B} \rangle \}= \frac{1}{i \hbar}
\langle [\hat{A},\hat{B}] \rangle\, , 
\end{equation}
extended to all moments by using linearity and the Leibniz rule.  The phase
space has boundaries according to Heisenberg's uncertainty relation
\begin{equation}
\Delta(q_j^2)\Delta(\pi_k^2)- \Delta(q_j \pi_k)^2 \geq \frac{\hbar^2}{4}
\delta_{jk} 
\end{equation}
and higher-order analogs.

Any given state (which may be pure or mixed) is therefore represented by a
point in phase space defined by the corresponding basic expectation values and
moments.  A state is considered semiclassical if its moments obey the
hierarchy
\begin{equation} \label{hierarchy}
 \Delta\left(q_1^{k_1}\cdots q_N^{k_N}\pi_1^{l_1}\cdots \pi_N^{l_N}\right)=
 O\left(\hbar^{\frac{1}{2}\sum_n (l_n+k_n)}\right)
\end{equation} 
which is satisfied, for instance, by a Gaussian, but includes also a more
general class of states. A semiclassical truncation of order $s$ of the
quantum system is defined as the submanifold spanned by the basic expectation
values and moments such that $\sum_n(l_n+k_n)\leq s$, which implies variables
up to order $\frac{1}{2}s$ in $\hbar$ according to the semiclassical
hierarchy. The Poisson bracket that results from (\ref{Poisson}) can
consistently be restricted to any semiclassical truncation by ignoring in
$\{\Delta_1,\Delta_2\}$ all terms of order higher than $s$ in moments.  In
this restriction, the product of a moment of order $s_1$ and a moment of order
$s_2$ is considered of semiclassical order $s_1+s_2$, while the product of a
moment of order $s_1$ with $\hbar^{s_2}$ is of order $s_1+2s_2$
\cite{Counting}. For given $s$, the Poisson tensor on the semiclassical
truncation of order $s$ is, in general, not invertible. Therefore,
semiclassical truncations and the resulting effective potentials  cannot be
formulated within symplectic geometry.

The Hamilton operator $\hat{H}$ determines a Hamilton function
$\langle\hat{H}\rangle$ on state space, which can be restricted to any
semiclassical truncation of order $s$ to define an effective Hamilton function
of semiclassical order $s$. We assume that each contribution to the Hamilton
operator is Weyl-ordered in basic operators. Any Hamilton operator that does
not obey this condition can be brought to Weyl-ordered form by using the
canonical commutation relations, which results in terms that explicitly depend
on $\hbar$. In order to compute an effective Hamiltonian of order $s$ for a
given Hamilton operator $H(\hat{q}_j,\hat{\pi}_k)$, we use
\begin{eqnarray} \label{Heff}
H_{{\rm eff},s}
&=& \langle H(\hat{q}_j + (\hat{q}_j - q_j), \hat{\pi}_k + (\hat{\pi}_k -
\pi_k))\rangle \\ 
&=& H(q_j,\pi_k)+ \sum \limits_{\sum_n(j_n+k_n)=2}^{s}
 \frac{\partial^n H(q,\pi)}{\partial q_1^{j_1}\cdots \partial
  q_N^{j_N} \partial \pi_1^{k_1}\cdots \partial \pi_N^{k_N}}
\frac{\Delta\left(q_1^{j_1}\cdots q_N^{j_N}\pi_1^{k_1}\cdots
    \pi^{k_N}\right)}{j_1!\cdots j_N! k_1!\cdots k_N!} \, .\nonumber
\end{eqnarray}
This expansion is reduced to a finite sum if $\hat{H}$ is polynomial in basic
operators, in which case the expansion serves the purpose of expressing the
expectation value of products of basic operators in terms of central
moments. For a non-polynomial Hamilton operator, the expansion is a formal
power series in $\hbar$.  The definition of our Poisson bracket ensures that
Hamilton's equations
\begin{eqnarray}
\dot{f}(\langle\cdot\rangle,\Delta) = \{ f(\langle\cdot\rangle,\Delta), H_{{\rm
  eff,s}}\}
\end{eqnarray}
on any semiclassical truncation are consistent with Heisenberg's equations of
motion evaluated in a state.

\subsection{Examples}

For a single pair of classical degrees of freedom, $N=1$, the phase space of
the semiclassical truncation of order two is five-dimensional (and therefore
cannot be symplectic). In addition to the basic expectation values $q$ and
$\pi$, there are two fluctuation variables, $\Delta(q^2)$ and $\Delta(\pi^2)$,
and the covariance $\Delta(q\pi)$. The non-zero Poisson brackets of these
variables are given by
\begin{eqnarray} \label{qpiDelta}
 \{q,\pi\} &=& 1\\
 \{\Delta(q^2),\Delta(q\pi)\} &=& 2\Delta(q^2)\\
 \{\Delta(q\pi),\Delta(\pi^2)\} &=& 2\Delta(\pi^2)\\
 \{\Delta(q^2),\Delta(\pi^2)\} &=& 4\Delta(q\pi)
\end{eqnarray}
which are linear and equivalent to the Lie algebra ${\rm sp}(2,{\mathbb R})$.

More generally, the second-order semiclassical truncation for $N$ pairs of
classical degrees of freedom is equivalent to ${\rm sp}(2N,{\mathbb R})$
\cite{Bosonize}. Third-order semiclassical truncations also have linear
Poisson brackets which are no longer semisimple: Within a higher-order
semiclassical truncation, the Poisson bracket of two third-order moments is
a sum of fourth-order moments and products of second-order moments, all of which
are of order four and set to zero in a third-order truncation. Moreover, the
Poisson bracket of a second-order moment and a third-order moment is
proportional to a third-order moment, for instance
\begin{equation}
 \{\Delta(q^2),\Delta(q^2\pi)\}= 2\Delta(q^3)\quad,\quad
 \{\Delta(q^2),\Delta(q\pi^2)\}= 4\Delta(q^2\pi)\quad,\quad
 \{\Delta(q^2),\Delta(\pi^3)\}=6\Delta(q\pi^2)
\end{equation}
for $N=1$.  The third-order moments in a semiclassical truncation of order
three therefore form an Abelian ideal, and the corresponding Lie algebra is
not semisimple. (For $N=1$, the Lie algebra is the semidirect product ${\rm
  sp}(2,{\mathbb R})\ltimes{\mathbb R}^4$ where ${\rm sp}(2,{\mathbb R})$ acts
according to its spin-$3/2$ representation \cite{Bosonize}.)

For orders higher than three, the Poisson brackets are non-linear and
therefore do not define Lie algebras. A general expression is given by
\cite{EffAc,HigherMoments}
\begin{eqnarray}\label{MomentBrackets}
\{\Delta(q^bp^a),\Delta(q^dp^c)\}&=&
a\, d \, \Delta(q^bp^{a-1}) \Delta(q^{d-1}p^c) - b c \Delta(q^{b-1}p^a)
\Delta(q^dp^{c-1})\nonumber\\
&&+\sum_{{\rm odd}\;n=1}^M
\left(\frac{i\hbar}{2}\right)^{n-1}
K_{abcd}^{n}\, \Delta(q^{b+d-n}p^{a+c-n})
\end{eqnarray}
where $M={\rm min}(a + c, b + d, a + b, c + d)$ and
\begin{equation}
K_{abcd}^n = \sum_{m= 0}^{n} (-1)^m m!(n-m)!
\left(\!\!\begin{array}{c}
a\\m
\end{array}\!\!\right)
\left(\!\!\begin{array}{c}
b\\n-m
\end{array}\!\!\right)
\left(\!\!\begin{array}{c}
c\\n-m
\end{array}\!\!\right)
\left(\!\!\begin{array}{c}
d\\m
\end{array}\!\!\right) \,.
\end{equation}
The inclusion of only odd $n$ in the sum ensures that all coefficients are
real. Terms containing $\Delta(q)$ or $\Delta(p)$ are considered zero: They
correspond to expectation values of the form $\langle\hat{a}-a\rangle=0$ which
are identically zero.

\subsection{Purity}
\label{s:Purity}

The collection of all moments determines a state, provided it obeys conditions
that follow from uncertainty relations. Since moments are defined using
expectation values, which can be computed from a pure or mixed state, they may
describe a pure or mixed state. In general, it is not easy to determine the
purity of a state described by moments without first reconstructing a density
matrix from them. As we will see, however, canonical variables for moments can
provide indications as to possible impurity parameters. In preparation of this
application, we discuss here ingredients for possible reconstructions of
states from a given set of moments.

If the state is pure, it is sufficient to consider only the moments
$\Delta(q^n)$ and $\Delta(q^{n-1}\pi)$ to reconstruct a wave function
\cite{EffAc}. For instance, we can use Hermite polynomials $H_n(q)$ and their
coefficients $h_{n,l}$ defined such that $H_n(q)=\sum_l h_{n,l}q^l$. The
expectation values $a_n = \langle \hat{q}^n\rangle$ can then be used to
compute
\begin{equation}
 c_n=\sum_l h_{n,l} a_l= \int{\rm d}q |\psi(q)|^2 H_n(q)\,,
\end{equation}
from which we obtain the probability density
\begin{equation}
 |\psi(q)|^2 = e^{-q^2} \sum_n \frac{c_n}{2^n\pi n!} H_n(q)
\end{equation}
using the orthonormality relation of Hermite polynomials. 

Using $b_n=\langle\hat{q}^n\hat{\pi}\rangle$, the phase $\alpha(q)$ of the
wave function $\psi(q)=\exp(i\alpha(q))|\psi(q)|$ then follows from
\begin{eqnarray}
 {\rm Re}b_n&=& {\rm Re} \int{\rm d}q \psi^* q^n \frac{\hbar}{i} \frac{{\rm
     d}\psi}{{\rm d}q}\\
&=& {\rm Re} \int{\rm d}q e^{-i\alpha}|\psi|
 q^n\frac{\hbar}{i} \left(i\frac{{\rm d}\alpha}{{\rm d}q} e^{i\alpha}|\psi|+
   e^{i\alpha} \frac{{\rm d}|\psi|}{{\rm d}q}\right)\\
&=& \hbar \int{\rm d}q |\psi|^2 q^n \frac{{\rm d}\alpha}{{\rm d}q}\,.
\end{eqnarray}
If we define 
\begin{equation}
 d_n = \sum_l h_{n,l} {\rm Re}b_n= \hbar \int{\rm d}q |\psi|^2 \frac{{\rm
     d}\alpha}{{\rm d}q} H_n(q)\,,
\end{equation}
we reconstruct
\begin{equation}
 \frac{{\rm d}\alpha}{{\rm d}q} = \frac{e^{-q^2}}{\hbar|\psi|^2} \sum_n
 \frac{d_n}{2^n\pi n!}H_n(q)\,.
\end{equation}
Integration gives $\alpha(q)$ up to an arbitrary constant phase.

In order to reconstruct a density matrix, we need all moments. First, position
moments are given by 
\begin{equation}
  \Delta(q^a)= {\rm tr}((\hat{q}-\langle\hat{q}\rangle)^a\hat{\rho})= \int
  (q-\langle\hat{q}\rangle)^a   \rho(q,q){\rm d}q 
\end{equation}
from which we can reconstruct the diagonal part $\rho(q,q)$ using orthogonal
polynomials. Using momentum-dependent moments, we can compute the values of
\begin{equation}
 {\rm tr}((\hat{q}-\langle\hat{q}\rangle)^a \hat{\pi}^b\hat{\rho})=
 \left(\frac{\hbar}{i}\right)^b \int
 (q-\langle\hat{q}\rangle)^a \left.\frac{\partial^b\rho(y,q)}{\partial
     y^b}\right|_{y=q} {\rm d}q
\end{equation}
and use them in
\begin{eqnarray}
 \sum_b \frac{1}{b!} \left(\frac{id}{\hbar}\right)^b
 {\rm tr}((\hat{q}-\langle\hat{q}\rangle)^a \hat{\pi}^b\hat{\rho})&=& \int
 (q-\langle\hat{q}\rangle)^a \sum_b \frac{d^b}{b!}
 \left.\frac{\partial^b\rho(y,q)}{\partial y^b}\right|_{y=q}{\rm d}q\nonumber\\
&=& \int
 (q-\langle\hat{q}\rangle)^a \rho(q+d,q){\rm d}q
\end{eqnarray}
to reconstruct $\rho(q+d,q)$ for arbitrary $q$ and $d$.

In a semiclassical truncation we have incomplete information about the moments
and it may be impossible to tell with certainty whether truncated moments
correspond to a pure or mixed state. However, if there are parameters that
appear only in moments of the form $\Delta(q^a\pi^b)$ with $b>1$, they may be
considered candidates for impurity parameters. We will see several examples in
our derivation of canonical variables for moments.

\subsection{Casimir--Darboux coordinates}

Since the brackets (\ref{MomentBrackets}) are non-canonical, it is not
possible to interpret the moments directly in terms of configuration variables
and momenta. However, the Darboux theorem and its generalization to Poisson
manifolds guarantees that one can always choose coordinates that are
canonical, together with a set of Casimir coordinates that have vanishing
Poisson brackets with all other variables. The required transformation from
moments to Casimir--Darboux variables of this form is, in general,
non-linear. In \cite{Bosonize}, we have developed a systematic method to
derive such transformations, based on a proof of Darboux' theorem given in
\cite{Arnold}. We have applied this method to semiclassical truncations in
\cite{Bosonize}, which we review here with further details in the relevant
integrations. 

\subsubsection{Single pair of degrees of freedom at second order}

We illustrate the method for the case of a semiclassical truncation of order
two for a single canonical pair of degrees of freedom. In this case,
Casimir--Darboux variables had already been found independently in
\cite{GaussianDyn,QHDTunneling}.

The relevant Poisson brackets of second-order moments are given in
(\ref{qpiDelta}). The procedure starts by choosing a function that plays the
role of the first canonical coordinate. It is convenient to have a quantum
fluctuation as one of the configuration variables, and therefore we choose
$s=\sqrt{\Delta(q^2)}$. This function, viewed formally as a Hamiltonian, is
the generator of a Hamiltonian flow on phase space defined by 
\begin{equation}\label{df}
 \frac{{\rm d}f(\Delta(q^2),\Delta(q\pi),\Delta(\pi^2)}{{\rm d}\epsilon}=
 \{f(\Delta(q^2),\Delta(q\pi),\Delta(\pi^2),s\}\,.
\end{equation}
If we already knew canonical coordinates, it would be obvious that the Poisson
bracket on the right-hand side of this equation changes only the variable
$p_s$ canonically conjugate to $s$, and therefore the derivative should be
equal to the (negative) partial derivative of $f$ by $p_s$. Since we do not
know $p_s$ yet, we revert this argument and implicitly define $p_s$ such that
the derivatives in (\ref{df}) equal the negative partial deirvative by $p_s$
for any function $f$. In particular, for the three second-order moments we
obtain
\begin{eqnarray}
 \frac{\partial\Delta(q^2)}{\partial p_s} &=&
 -\{\Delta(q^2),\sqrt{\Delta(q^2)}\}=0\\ 
 \frac{\partial\Delta(q \pi)}{\partial p_s} &=&
 -\{\Delta(q\pi),\sqrt{\Delta(q^2)}\}=  \sqrt{\Delta(q^2)}=s \label{Diff1}\\
 \frac{\partial\Delta(\pi^2)}{\partial p_s} &=&
 -\{\Delta(\pi^2),\sqrt{\Delta(q^2)}\}=2\frac{\Delta(q
   \pi)}{\sqrt{\Delta(q^2)}}= 2\frac{\Delta(q \pi)}{s}\,.  \label{Diff2}
\end{eqnarray}
By construction, these are partial differential equations in which $s$ is held
constant. We can easily solve (\ref{Diff1}) by
\begin{equation} \label{Deltaqp}
 \Delta(q \pi)=s p_s +f_1(s)
\end{equation}
with a free function $f_1$ depending only on $s$.
Inserting this solution in (\ref{Diff2}), we have
\begin{equation}\label{Deltap}
 \Delta(\pi^2)=p_s^2+2\frac{f_1(s)}{s} p_s +f_2(s)
\end{equation}
with another free function $f_2$ depending only on $s$.

Computing $\{\Delta(q \pi),\Delta(\pi^2)\}$ using the canonical nature of the
variables $s$ and $p_s$, and requiring that it equal $2\Delta(\pi^2)$ implies
two equations:
\begin{equation}
 \frac{{\rm d}f_1}{{\rm d}s}=\frac{f_1}{s} \quad,\quad \frac{{\rm d}f_2}{{\rm
     d}s} = 2\frac{f_1}{s^2} \frac{{\rm d}f_1}{{\rm d}s} - 2\frac{f_2}{s} \,.
\end{equation}
They are solved by
\begin{equation} \label{fU}
 f_1(s)=U_2s \quad,\quad f_2(s)=\frac{U_1}{s^2}+U_2^2
\end{equation}
with constants $U_1$ and $U_2$. We can eliminate $U_2$ by a canonical
transformation replacing $p_s$ with $p_s+U_2$. The constant $U_1$ is the
Casimir coordinate. The resulting moments in terms of Casimir--Darboux
variables are
\begin{equation} \label{sps}
 \Delta(q^2) = s^2\quad,\quad
\Delta(q \pi) = s  p_s\quad,\quad
\Delta(\pi^2) = p_s^2+\frac{U_1}{s^2}
\end{equation}
as in \cite{GaussianDyn,QHDTunneling}.  

In general, it may be difficult to recognize a variable such as $U_1$ as a
Casimir coordinate. In such a case, the flow 
generated by $s$ or $s^2=\Delta(q^2)$ is again useful:
\begin{equation}
 \frac{{\rm d}\Delta(q\pi)}{{\rm d}\epsilon} = -2\Delta(q^2)
 \quad,\quad \frac{{\rm d}\Delta(\pi^2)}{{\rm d}\epsilon} = -4\Delta(q\pi)\,.
\end{equation}
The solutions are similar to what we already used,
$\Delta(q\pi)[\epsilon]=-2\Delta(q^2)\epsilon+d$ for the first equation and 
$\Delta(\pi^2)[\epsilon]=4\Delta(q^2)\epsilon^2-4d\epsilon+e$ for the second
equation, with constants $d$ and $e$.
But now we use these equations to eliminate $\epsilon$ instead of solving for
$p_s$. Inserting $\epsilon=\frac{1}{2}(d-\Delta(q\pi)[\epsilon])/s^2$ in
$\Delta(\pi^2)[\epsilon]$ implies 
\begin{equation}
 \Delta(\pi^2)[\epsilon]= \frac{\Delta(q\pi)[\epsilon]^2}{\Delta(q^2)}-
 3\frac{d^2}{\Delta(q^2)}+e\,.
\end{equation}
The combination
$U_1=\Delta(q^2)\Delta(\pi^2)[\epsilon]-\Delta(q\pi)[\epsilon]^2= -3d^2+es^2$
is therefore independent of $\epsilon$. Since ${\rm d} U_1/{\rm
  d}\epsilon=\{U_1,\Delta(q^2)\}=0$, $U_1$ is a coordinate Poisson orthogonal
to $s$. It is also Poisson orthogonal to $p_s$ by construction, and therefore
represents the Casimir variable of this system.

\subsubsection{Single pair of degrees of freedom at third order}

We now try to find an extension of our Casimir--Darboux coordinates to third
order. There are now seven moments, and the rank of the Poisson tensor shows
that there is a single Casimir variable. We must therefore derive two
additional pairs of canonical degrees of freedom. Since Darboux coordinates
are defined only up to canonical transformations, the form in which they
appear in the moments is not unique and subject to choices. For now, we make a
choice motivated by the canonical form we just derived at second order: We
assume that $\Delta(q^2)$ depends only on one of the new canonical pairs,
\begin{equation}
\Delta(q^2)=s_1^2
\end{equation}
from which it quickly follows, by a calculation similar to our second-order
example, that
\begin{equation}
\Delta(q \pi)= s_1 \, p_{1} 
\end{equation}
is a consistent (but not unique) choice of introducing the first momentum. 

The remaining canonical pairs must be such that they have zero Poisson
brackets with $s_1$ and $p_1$, or with $\Delta(q^2)$ and $\Delta(q\pi)$
according to our first choices. The same procedure that we used to derive
$U_1$ as a coordinate Poisson orthogonal to both $s$ and $p_s$ at second order
can also be used here, but now we have five additional moments which should be
expressed in terms of functions Poisson orthogonal to $s$ and $p_s$. By
systematically computing the flows of all the remaining moments generated by
$s_1$ and $p_1$ and eliminating flow parameters, it follows that the following
functions of moments are Poisson orthogonal to $s_1$ and $p_1$:
\begin{eqnarray*}
\label{f}
f_1&=&\Delta(q^2)\Delta(\pi^2)-\Delta(q\pi)^2\\
f_2&=&\Delta(q^2)\frac{\Delta(q^2\pi)}{\Delta(q^3)}-\Delta(q\pi)\\
f_3&=&\frac{\Delta(q^2)^2}{\Delta(q^3)^2}\left(\Delta(q^2\pi)^2-
  \Delta(q\pi^2)\Delta(q^3) \right)\\
f_4 &=& 2 \Delta(q\pi)+\Delta(q^2)\frac{\Delta(q^3)\Delta(\pi^3)-
  \Delta(q\pi^2)\Delta(q^2\pi)}{\Delta(q^2\pi)^2-\Delta(q\pi^2)\Delta(q^3)} \, .
\end{eqnarray*}
One additional variable can be derived independently from the Casimir function
of the Lie algebra that corresponds to third-order moments,
\begin{eqnarray} \label{f5}
f_5:=U_1^4&=&\left(\Delta(q^2\pi)\Delta(q\pi^2)-
  \Delta(q^3)\Delta(\pi^3)\right)^2\\ 
&&-4\left(\Delta(q\pi^2)^2-\Delta(q^2\pi)\Delta(\pi^3)\right)
\left(\Delta(q^2\pi)^2-\Delta(q^3)\Delta(q\pi^2)\right)\, .
\end{eqnarray}
(The fourth power of $U_1$ is chosen such that $U_1$ is of third order just
like the moment order considered here.)  While $f_5$ Poisson commutes with all
other $f_i$, $(f_1,f_2,f_3,f_4)$ have non-linear brackets
\begin{eqnarray}
\left\{f_1,f_2\right\}&=& 2 f_1+2 f_2^2+4 f_3\label{f1f2}\\
\left\{f_1,f_3\right\}&=& 12 f_2 f_3+2 f_3 f_4\\
\left\{f_1,f_4\right\}&=&-4 f_1-f_4^2+4 f_3-\left(2 f_2+f_4\right)^2\\
\left\{f_2,f_3\right\}&=&  -4 f_3\\
\left\{f_2,f_4\right\}&=& -4 f_2-2 f_4\\
\left\{f_3,f_4\right\}&=& -8 f_3 \label{f3f4}
\end{eqnarray}
with one another.

We are now ready to choose our second configuration variable. We define
\begin{equation}
s_2=f_3 \, ,
\end{equation}
such that
\begin{equation}
\frac{\partial f_1}{\partial p_2}=-12 s_2 f_2-2 s_2 f_4\quad,\quad
\frac{\partial f_2}{\partial p_2}=4 s_2\quad,\quad
\frac{\partial f_4}{\partial p_2}=-8 s_2 
\end{equation}
can be used to determine the second momentum variable.
Integrating the last two equations and inserting the results in the first one
gives 
\begin{eqnarray} 
f_1&=&-16 s_2^2p_2^2-s_2\left(12 g_2+2 g_4\right)p_2+g_1 \label{f1}\\
f_2&=&4 s_2 p_2+g_2\label{f2}\\
f_4&=&-8 s_2 p_2+g_4  \label{f4}
\end{eqnarray}
with three functions $g_1$, $g_2$ and $g_4$ independent of $p_2$. (They can
therefore depend on $s_2$ and the remaining canonical pair, $s_3$ and
$p_3$, as well as the Casimir variable $U_1$.) Since we are interested in
deriving $p_2$, we can choose the free functions such that it is easy to
invert (\ref{f1}), (\ref{f2}) or (\ref{f4}) for $p_2$. A wrong choice at this
point could result in a degenerate system that does not allow us to derive
all canonical pairs. Since we know how many canonical pairs we obtain, a
little bit of trial and error quickly shows when a choice is suitable.  If we
choose $g_4=-6 g_2$, we obtain
\begin{equation} \label{p2}
p_2=\frac{6 f_2+f_4}{16 s_2}
\end{equation}
from a combination of (\ref{f2}) and (\ref{f4}), as well as
\begin{equation} 
g_1=f_1+\frac{\left(6 f_2+f_4\right)^2}{16}\quad,\quad
g_2=-\frac{1}{2}f_2-\frac{1}{4}f_4 \, .
\end{equation}

By construction, $g_1$ and $g_2$ do not depend on $p_2$, but we have not made
sure yet that they do not depend on $s_2$ either. Since $s_2$ is defined as
$s_3$, the Poisson brackets (\ref{f1f2})--(\ref{f3f4}) can be used to show
that $g_1$ and $g_2$ do, in fact, depend on $s_2$. The same Poisson brackets
determine the canonical flow generated by $p_2$ in (\ref{p2}) on $g_1$ and
$g_2$. By eliminating the flow parameter as in some of the previous steps, we
find that  the combinations
\begin{eqnarray}
p_3&=&\frac{g_2}{\sqrt{s_2}}\\
s_3&=&\frac{2 g_1-7 s_2+10 p_3^2 s_2}{6 \sqrt{s_2}(4 p_3^2-1)}
\end{eqnarray}
are independent of $s_2$ and are therefore Poisson orthogonal to all previously
constructed canonical pairs. They determine our final pair $(s_3,p_3)$.

In order to express moments in terms of canonical pairs and the Casimir
variable, we insert the functions
\begin{eqnarray}
f_1 &=& 3 \sqrt{s_2}\left(-1+4 p_3^2\right)s_3+\frac{1}{2}\left(7 s_2-10 s_2
  p_3^2\right)-16 s_2^2 p_2^2\\ 
f_2 &=& \sqrt{s_2}p_3+4 s_2 p_2\\
f_3 &=& s_2\\
f_4 &=& -4 \sqrt{s_2}p_3-8 s_2 p_2\\
f_5 &=& U_1 ^4
\end{eqnarray}
in (\ref{f}) and (\ref{f5}) and invert the resulting relations for
\begin{eqnarray}
\Delta(q^2)&=&s_1^2\quad,\quad \Delta(q\pi)= s_1 p_1\\
\Delta(\pi^2)&=&p_1^2+\frac{f_1}{s_1^2}= p_1^2+ \frac{3 \sqrt{s_2}\left(4
    p_3^2-1\right)s_3+\frac{1}{2}s_2\left(7 -10  
  p_3^2\right)-16 s_2^2 p_2^2}{s_1^2} \label{Deltapi1}\\
\Delta(\pi^3)&=&
\frac{U_1}{s_1^3\sqrt{2s_2^{3/2}\sqrt{1-4p_3^2}}}\Phi(s_i,p_i)\\ 
\Delta(q\pi^2)&=&\frac{U_1}{s_1 \sqrt{2s_2^{3/2}\sqrt{1-4p_3^2}}}\left(p_1 s_1
  +\left(p_3-1\right)\sqrt{s_2}+4 s_2 p_2\right)\\ 
&&\times\left(p_1 s_1 +\left(1+p_3\right)\sqrt{s_2}+4 s_2
  p_2\right)\\ 
\Delta(q^2\pi)&=&\frac{U_1}{\sqrt{2s_2^{3/2}\sqrt{1-4p_3^2}}}\left(p_1
  s_1^2+s_1\left(p_3 \sqrt{s_2}+4 s_2 
    p_2\right)\right)\\ 
\Delta(q^3)&=&\frac{U_1s_1^3}{\sqrt{2s_2^{3/2}\sqrt{1-4p_3^2}}}\, , 
\end{eqnarray}
where
\begin{eqnarray}
\Phi(s_i,p_i)&=&p_1^3 s_1^3+3 p_1^2 p_3 s_1^2 \sqrt{s_2}
+ 3 p_1 s_1 s_2 \left(-1+p_3^2+4 p_1 s_1 p_2\right)+64 p_2^3 s_2^3\\
&&+p_3 s_2^{3/2}\left(-7+p_3^2+24 p_1 p_2 s_1\right)+48 p_3 p_2^2 s_2^{5/2}
+12 p_2 s_2^2\left(-1+p_3^2+4 p_1 s_1 p_2\right)\,.\nonumber
\end{eqnarray}
More compactly, some of the momentum-dependent moments can be written as
\begin{eqnarray}
 \Delta(\pi^3)&=&
\frac{U_1\left(P^3-3P-4p_3s_2^{3/2}\right)}{s_1^3\sqrt{2s_2^{3/2}\sqrt{1-4p_3^2}}}
\\
\Delta(q\pi^2)&=&\frac{U_1\left(P^2-s_2\right)}{s_1
  \sqrt{2s_2^{3/2}\sqrt{1-4p_3^2}}}  \\
\Delta(q^2\pi)&=&\frac{U_1s_1P}{\sqrt{2s_2^{3/2}\sqrt{1-4p_3^2}}} 
\end{eqnarray}
if we introduce $P=p_1s_1+p_3\sqrt{s_2}+ 4s_2p_2$. Note that $s_3$ does not
appear in any $\Delta(q^a\pi^b)$ with $b\leq 1$, and may therefore be a
candidate for the impurity of a state.

\subsubsection{Third order by ansatz}

As we have seen, several choices have to be made in the process of deriving
Casimir--Darboux coordinates. Some choices may lead to degenerate systems in
which a smaller number of canonical pairs results, and which should therefore
be discarded. However, even within the class of non-degenerate systems, there
cannot be a unique set of Casimir--Darboux coordinates because one can always
apply canonical transformations of Darboux variables. Depending on the
application, some choices may lead to more useful realizations of canonical
variables than others. Staying with the third-order system for a single pair
of canonical degrees of freedom, we now apply an alternative method which
works by ansatz and therefore is somewhat less systematic than the previous
procedure. However, it makes it easier to implement certain properties such as
a simplified version of $\Delta(\pi^2)$ in (\ref{Deltapi1}) with
$s_i$-independent coefficients. As we will see, such a version greatly
simplifies the effective dynamics, but it does not always exist, in particular
if we have more than one pair of classical degrees of freedom.

We make the ansatz
\begin{eqnarray} \label{ansatz}
\Delta(\pi^2) =\sum_{i=1}^{3}p_i^2+F(s_1,s_2,s_3)\quad&,&\quad \Delta(q \pi)
=\sum_{i=1}^{3}s_i p_i\\ 
\Delta(q^2) =\sum_{i=1}^{3}s_i^2\quad&,&\quad
\Delta(q^3) =\sum_{i=1}^{3}s_i^3 \label{ansatz2}
\end{eqnarray}
introducing three canonical pairs, as required.  The function
$F(s_1,s_2,s_3)$, which is assumed to be independent of the momenta, is
subject to consistency conditions that follow from the required Poisson
brackets of moments.  Once we have a consistent $F$, we can generate all the
remaining moments by taking successive Poisson brackets with $\Delta(\pi^2)$:
\begin{equation}\label{Generate}
 \Delta(q^{m-1}\pi^{n+1})=-\frac{1}{2m}
 \left\{\Delta(\pi^2),\Delta(q^m\pi^n)\right\}\,, 
\end{equation}
starting with $m=3$, $n=0$ in which case we have defined $\Delta(q^3)$ in
(\ref{ansatz2}) and can derive 
\begin{eqnarray}
 \Delta(q^2\pi) &=& \sum_i p_is_i^2\\
 \Delta(q\pi^2) &=& \sum_i p_i^2s_i- \frac{1}{4} \sum_is_i^2\frac{\partial
   F}{\partial s_i}\\
 \Delta(\pi^3)&=& \sum_ip_i^3- \frac{1}{4} \sum_ip_i\left(4s_i\frac{\partial
     F}{\partial s_i}+ \sum_j s_j^2\frac{\partial^2F}{\partial s_i\partial
     s_j}\right)\,.
\end{eqnarray}
Since we have explicitly used all three canonical pairs expected for a
third-order truncation, $F$ depends on one further parameter, $U$, which will
be the Casimir coordinate. Since $F$ and therefore $U$ appear only in moments
which have at least two momentum factors, $U$ is a candidate for an impurity
parameter in this mapping.

Equation~(\ref{Generate}) also applies to second-order moments, $m+n=2$. Since
we have defined all three second-order moments in (\ref{ansatz}), we obtain
consistency conditions on $F$. 
We first compute
\begin{equation}
 \{\Delta(\pi^2),\Delta(q^2)\}= -4\sum_is_ip_i
\end{equation}
and from this
\begin{equation}
 \{\Delta{\pi^2},\{\Delta(\pi^2),\Delta(q^2)\}\}= 8\sum_ip_i^2- 4\sum_i
 s_i\frac{\partial F}{\partial s_i}\,.
\end{equation}
The condition 
\begin{equation}
 \{\Delta{\pi^2},\{\Delta(\pi^2),\Delta(q^2)\}\}= 8\Delta(\pi^2)
\end{equation}
then implies 
\begin{equation}\label{con1}
 \sum_{i}s_i \frac{\partial F}{\partial s_i}=-2F
\end{equation}
and therefore $F$ is homogeneous of degree $-2$ if all $s_i$ are rescaled by
the same constant.

Applying further Poisson brackets with $\Delta(\pi^2)$ does not give new
conditions. For instance, 
\begin{equation}
 0=\left\{\Delta(\pi^2),
\left\{\Delta(\pi^2),\left\{\Delta(\pi^2),\Delta(q^2)
\right\}\right\}\right\} 
\end{equation}
is equivalent to 
\begin{equation}
 0= 8\left(3\sum_ip_i\frac{\partial F}{\partial s_i}+
 \sum_{i,j}p_is_j\frac{\partial^2 F}{\partial s_i\partial s_j}\right)\,.
\end{equation}
Since the $s_i$ and $p_i$ can be varied independently, the condition
implies that all three $\partial F/\partial s_i$ are homogeneous of degree
$-3$ if all $s_i$ are rescaled by
the same constant, which follows from $F$ being of degree $-2$.

Another consistency condition can be derived by looking at the third order
moments:
\begin{equation}
 0=\left\{\Delta(\pi^2),\left\{\Delta(\pi^2),
\left\{\Delta(\pi^2),\left\{\Delta(\pi^2),\Delta(q^3)
\right\}\right\}\right\} \right\}
\end{equation}
is equivalent to 
\begin{eqnarray}\label{con2}
0&=&6 \sum_{i}p_i^2 \frac{\partial F}{\partial s_i}+4 \sum_{ij} p_i s_i p_j
\frac{\partial^2 F}{\partial s_i \partial s_j}+\frac{1}{2}\sum_{ijk}s_i^2 p_j
p_k \frac{\partial^3 F}{\partial s_i \partial s_j \partial s_k}\\ 
&&- \frac{3}{2}\sum_{i}s_i \left(\frac{\partial F}{\partial
    s_i}\right)^2-\frac{1}{4}\sum_{ij}s_i^2 \frac{\partial F}{\partial s_j}
\frac{\partial^2 F}{\partial s_i \partial s_j} \,.
\end{eqnarray}
This condition is generally independent from ({\ref{con1}}).  For example, the
solution $F=\sum_i U/s_i^2$ of (\ref{con1}) is not a solution of (\ref{con2}).

One further condition has to be imposed, which is the invertibility of the
mapping from moments to $(s_i,p_{s_j})$. (Otherwise one could choose the
trivial solution $F=0$.) For any given $F$, this condition can be checked by
computing the Jacobian of the transformation, and it is fulfilled, for
instance, by the solutions
\begin{equation} \label{FU}
F(s_1,s_2,s_3)=\sum_{i<j}\frac{U}{(s_i-s_j)^2}
\end{equation}
of (\ref{con1}) and (\ref{con2}), where $U$ is the Casimir variable.
Therefore, there is a faithful mapping from moments to canonical coordinates
at the third order, such that moments are quadratic in the new momenta with
$s$-independent coefficients. The ansatz used here provides a simplified
procedure to compute Casimir--Darboux coordinates, but only if moments
quadratic in momenta exist. The choice (\ref{FU}) is not unique, but it is
interesting because for $U>0$ it implies repulsive potentials between the
$s_i$ in an effective potential.

At this point, we have obtained two different canonical systems for the
third-order semiclassical truncation of a single classical degree of freedom,
with Casimir variables $U_1$ and $U$, respectively.  However, a direct
comparison of these two versions of the Casimir variable is difficult because
the two Poisson algebras we have canonically realized, in fact, differ from
each other in a subtle way: For the mapping derived with the ansatz we have
Poisson brackets of  third order moments of the form
$\left\{\Delta^3_i,\Delta^3_j\right\}=\mathcal{O}(\hbar^2)$. The right-hand
side is considered zero in a third-order semiclassical truncation, which
corresponds to an $\hbar$-order of $3/2$. For the mapping derived
systematically, however, we were able to exactly impose
$\left\{\Delta^3_i,\Delta^3_j\right\}=0$. Therefore, the two Casimir variables
are likely to differ from each other by terms of the order $\hbar^2$. 

Nevertheless, it is instructive to compute the Poisson bracket of the moments
derived with the ansatz with the Casimir $U_1$ that was derived
systematically. Assuming that $s$ and $p$ are of the order
$\mathcal{O}(\sqrt{\hbar})$ in a semiclassical state, computer algebra shows
that the Taylor expansion of the Poisson brackets $\left\{\Delta_{\rm
    ansatz},U_1(\Delta_{\rm ansatz})\right\}=\mathcal{O}(\hbar^{5/2})$ in
$\sqrt{\hbar}$ is zero within the third-order truncation. Therefore, the
Casimir variable derived systematically is a Casimir variable also for the
realization derived using an ansatz, up to a truncation error.

\subsubsection{Fourth order}

The solution at the third order can be extended in a rather direct manner to
the fourth order. Inspection of the rank of the Poisson tensor at this order
shows that we expect five canonical pairs of quantum degrees of freedom and
two Casimir variables. We then try the ansatz
\begin{eqnarray}
\Delta(\pi^2)&=&\sum_{i=1}^{5} p_i^2+\sum_{i>j}\frac{U}{(s_i-s_j)^2}\\
\Delta(q^2)&=&\sum_{i}s_i^2\\
\Delta(q^3)&=&C\sum_{i}s_i^3\,.
\end{eqnarray}
In addition to an extension of the third-order ansatz to five pairs of
canonical degrees of freedom, we have inserted a new parameter $C$ which
will play the role of the second Casimir variable. 

The moment $\Delta(q^4)$ can be generated from the Poisson bracket
$\left\{\Delta(\pi q^2),\Delta(q^3)\right\}=3 \Delta(q^2)^2-3
\Delta(q^4)$:
\begin{equation}
\Delta(q^4)=C^2 \sum_{i}s_i^4+\sum_{i,j} s_{i}^2 s_{j}^2
\end{equation}
We also need to check that the Poisson bracket is consistent at this
order. For instance, while an expansion of the right-hand side of 
\begin{equation}\label{con3}
0=\left\{\Delta(\pi^2),\left\{\Delta(\pi^2),\left\{\Delta(\pi^2),
\left\{\Delta(\pi^2),\left\{\Delta(\pi^2),\Delta(q^4)
\right\}\right\}\right\}\right\}\right\} \,,
\end{equation}
would be too complex to be shown here, computer algebra confirms that
(\ref{con3}) is indeed satisfied for our ansatz.  This result supports the
physical principle that (when $U>0$) the quantum coordinates feel a repulsive
potential between one another that goes as one over the square of the distance
between them.

\subsubsection{Second-order truncation for two pairs of classical
  degrees of freedom}

For two pairs of classical degrees of freedom, we have a ten-dimensional
submanifold of second-order moments. The Poisson tensor has rank eight, so
that we have to construct four canonical pairs and two Casimir variables.

\paragraph{First step:}
The system contains two subalgebras that correspond to a single degree of
freedom, given by
$\langle\Delta(q_1^2),\Delta(q_1\pi_1),\Delta(\pi_1^2)\rangle$ and
$\langle\Delta(q_2^2),\Delta(q_2\pi_2),\Delta(\pi_2^2)\rangle$. We can
therefore make use of some of our previous derivations if we choose the first
two configuration variables as $s_1=\sqrt{\Delta(q_1^2)}$ and
$s_2=\sqrt{\Delta(q_2^2)}$. We obtain solutions similar to (\ref{Deltaqp}) and
(\ref{Deltap}) with (\ref{fU}), but now the free functions $f_{q_1\pi_1}$,
$f_{\pi_1^2}$, $f_{q_2\pi_2}$ and $f_{\pi_2^2}$ in
\begin{equation} \label{Deltaqp1}
 \Delta(q_1\pi_1) = s_1p_1+f_{q_1\pi_1} \quad,\quad \Delta(\pi_1^2) =
 p_1^2+2\frac{p_1}{s_1}f_{q_1\pi_1}+ f_{q_1\pi_1}^2+\frac{f_{\pi_1^2}}{s_1^2}
\end{equation}
and
\begin{equation}\label{Deltaqp2}
 \Delta(q_2\pi_2) = s_2p_2+f_{q_2\pi_2} \quad,\quad \Delta(\pi_2^2) =
 p_2^2+2\frac{p_2}{s_2}f_{q_2\pi_2}+ f_{q_2\pi_2}^2+\frac{f_{\pi_2^2}}{s_2^2}
\end{equation}
may still depend on the remaining two canonical pairs, as well
as the two Casimirs.

Since $f_{q_1\pi_1}$, $f_{\pi_1^2}$, $f_{q_2\pi_2}$ and $f_{\pi_2^2}$ do not
depend on $s_1$, $p_1$, $s_2$ and $p_2$ by construction, they parameterize
coordinate Poisson orthogonal to the first two canonical pairs. However, it is
convenient to choose $f_{q_1\pi_1}=0=f_{q_2\pi_2}$ because the condition of
being Poisson orthogonal to $s_1$, $p_1$, $s_2$ and $p_2$ is then equivalent
to having vanishing Poisson brackets with the basic moments
$\Delta(q_1^2)=s_1^2$, $\Delta(q_1\pi_1)=s_1p_1$, $\Delta(q_2^2)=s_2^2$ and
$\Delta(q_2\pi_2)=s_2p_2$.  This leaves two functions,
\begin{equation}
 f_{\pi_1^2}=s_1^2\Delta(\pi_1^2)-s_1^2p_1^2=
 \Delta(q_1^2)\Delta(\pi_1^2)-\Delta(q_1\pi_1)^2 =:f_1
\end{equation}
and
\begin{equation}
 f_{\pi_2^2}=s_2^2\Delta(\pi_2^2)-s_2^2p_2^2=
 \Delta(q_2^2)\Delta(\pi_2^2)-\Delta(q_2\pi_2)^2=:f_2\,,
\end{equation}
out of the original free functions in (\ref{Deltaqp1}) and (\ref{Deltaqp2}),
which we can easily write in terms of moments.

In addition to $f_1$ and $f_2$, we need four further functions that Poisson
commute with the first two canonical pairs, or with $\Delta(q_1^2)$,
$\Delta(q_1\pi_1)$, $\Delta(q_2^2)$ and $\Delta(q_2\pi_2)$.  As before, we
find such variables by considering the flows generated by $\Delta(q_1^2)$,
$\Delta(q_1\pi_1)$, $\Delta(q_2^2)$ and $\Delta(q_2\pi_2)$. For instance, for
$\Delta(q_1\pi_1)$, the flows ${\rm d}/{\rm
  d}\epsilon=\{\cdot,\Delta(q_1\pi_1)\}$ on the remaining moments are
\begin{eqnarray}
 \frac{{\rm d}\Delta(q_1q_2)}{{\rm d}\epsilon} = \Delta(q_1q_2)\quad&,&\quad 
\frac{{\rm d}\Delta(q_1\pi_2)}{{\rm d}\epsilon} = \Delta(q_1\pi_2)\quad,\quad 
\frac{{\rm d}\Delta(q_2\pi_1)}{{\rm d}\epsilon} = -\Delta(q_2\pi_1)\quad,\quad
\nonumber\\ 
\frac{{\rm d}\Delta(\pi_1\pi_2)}{{\rm d}\epsilon} =-
\Delta(\pi_1\pi_2)\quad&,&\quad  
\frac{{\rm d}\Delta(q_1^2)}{{\rm d}\epsilon} = 2\Delta(q_1^2) \,.
\end{eqnarray}
These linear differential equations can easily be solved by
\begin{eqnarray}
 \Delta(q_1q_2)=c_1e^{\epsilon}\quad&,&\quad
 \Delta(q_1\pi_2)=c_2e^{\epsilon}\quad,\quad
 \Delta(q_2\pi_1)=c_3e^{-\epsilon}\quad,\quad\nonumber\\
 \Delta(\pi_1\pi_2)=c_4e^{-\epsilon}\quad&,&\quad
 \Delta(q_1^2)=c_5e^{2\epsilon}\,.
\end{eqnarray}
By eliminating $\epsilon$, we find that $\Delta(q_1q_2)\Delta(q_2\pi_1)$,
$\Delta(q_1q_2)\Delta(\pi_1\pi_2)$, $\Delta(q_1\pi_2)\Delta(q_2\pi_1)$,
$\Delta(q_1\pi_2)\Delta(\pi_1\pi_2)$ and
$\Delta(q_1^2)\Delta(\pi_1\pi_2)\Delta(q_2\pi_1)$ Poisson commute with
$\Delta(q_1\pi_1)$. However, these combinations are not necessarily invariant
under the flows generated by $\Delta(q_1^2)$, $\Delta(q_2^2)$ and
$\Delta(q_2\pi_2)$. After computing variables invariant with respect to any
one of these four flows, we find that the combinations
\begin{eqnarray}
 f_3 &=& \Delta(q_1\pi_2)\Delta(q_2\pi_1)-\Delta(q_1q_2)\Delta(\pi_1\pi_2)\\
 f_4 &=&
 \Delta(q_1^2)\frac{\Delta(q_2\pi_1)}{\Delta(q_1q_2)}-\Delta(q_1\pi_1)\\
 f_5 &=&
 \Delta(q_2^2)\frac{\Delta(q_1\pi_2)}{\Delta(q_1q_2)}-\Delta(q_2\pi_2)\\
 f_6 &=& \frac{\Delta(q_1^2)\Delta(q_2^2)}{\Delta(q_1q_2)^2}\,,
\end{eqnarray}
in addition to $f_1$ and $f_2$, are Poisson orthogonal to $s_1$, $p_1$, $s_2$
and $p_2$. Moreover, 
their mutual Poisson brackets are closed,
\begin{eqnarray}
 \{f_1,f_2\}&=&0= \{f_1,f_3\}=\{f_2,f_3\}\\
 \{f_1,f_4\} &=& 2(f_1+f_4^2)\quad,\quad
\{f_1,f_5\} = 2f_3f_6\quad,\quad \{f_1,f_6\} = 4f_4f_6\\
\{f_2,f_4\} &=& 2f_3f_6\quad,\quad \{f_2,f_5\} = 2(f_2+f_5^2)\quad,\quad
\{f_2,f_6\} = 4f_5f_6\\
\{f_3,f_4\} &=& f_1+f_3f_6+f_4^2\quad,\quad \{f_3,f_5\} =
f_2+f_3f_6+f_5^2\quad,\quad \{f_3,f_6\} = 2(f_4+f_5)f_6\\
\{f_4,f_5\} &=& (f_5-f_4)f_6\quad,\quad \{f_4,f_6\} = -2f_6(1-f_6)=\{f_5,f_6\}
\end{eqnarray}
and therefore form a Poisson manifold on which we can iterate our procedure,
expressing the $f_1$ in terms of further Casimir--Darboux variables.

\paragraph{Second step:}
We now define $s_3=f_6$, equal to the inverse of the
correlation between the two positions. It generates a flow to be identified
with the negative partial derivative with respect tp $p_3$, 
\begin{eqnarray}
 \frac{\partial f_1}{\partial p_3} &=& -\{f_1,f_6\} = -4s_3f_4\\
 \frac{\partial f_2}{\partial p_3} &=& -4s_3f_5\quad,\quad
 \frac{\partial f_3}{\partial p_3} = -2s_3(f_4+f_5)\\
 \frac{\partial f_4}{\partial p_3} &=& 2s_3(1-s_3)\quad,\quad
 \frac{\partial f_5}{\partial p_3} = 2s_3(1-s_3)\,.
\end{eqnarray}
The last two equations are solved by
\begin{equation} \label{f4f5}
 f_4=2s_3p_3(1-s_3)+g_4 \quad\mbox{and}\quad f_5=2s_3p_3(1-s_3)+g_5\,,
\end{equation}
after which the remaining equations can be solved by
\begin{eqnarray}
 f_1 &=& -4s_3^2(1-s_3)p_3^2- 4s_3p_3 g_4+g_1\\
 f_2 &=& -4s_3^2(1-s_3)p_3^2- 4s_3p_3 g_5+g_2\\
 f_3 &=& -4s_3^2(1-s_3)p_3^2- 2 s_3p_3 (g_4+g_5)+g_3\,.
\end{eqnarray}
The functions $g_i$ are independent of $p_3$.

As before, a choice is required to proceed because we have five free functions
$g_i$ but only one more canonical pair and two Casimir variables.  The choice
$g_5=-g_4$ simplifies $f_3$ and eliminates these functions from $f_4+f_5$
according to (\ref{f4f5}) and we obtain our third momentum
\begin{equation} \label{p3}
 p_3=\frac{f_4+f_5}{4s_3(1-s_3)}\,.
\end{equation}

We are left with four functions $g_1,\ldots,g_4$ which, by construction, are
independent of $p_3$. But they may depend on $s_3$ and
are therefore not Poisson orthogonal to the third canonical pair.
In order to  find combinations which Poisson commute
with $p_3$, we consider the flow generated by $f_4+f_5=4s_3(1-s_3)p_3$. From
\begin{eqnarray}
 g_1 &=& f_1+\frac{(f_4+f_5)^2}{4(1-f_6)}+
 \frac{1}{2}\frac{(f_4+f_5)(f_4-f_5)}{1-f_6}\\
 g_2 &=& f_2+\frac{(f_4+f_5)^2}{4(1-f_6)}-
 \frac{1}{2}\frac{(f_4+f_5)(f_4-f_5)}{1-f_6}\\
 g_3 &=& f_3+\frac{(f_4+f_5)^2}{4(1-f_6)}\\
g_4 &=& \frac{1}{2}(f_4-f_5)\,,
\end{eqnarray}
We obtain the brackets
\begin{eqnarray}
 \{g_1,f_4+f_5\} &=& 2(g_1+s_3g_3+g_4^2)\\
\{g_2,f_4+f_5\} &=& 2(g_2+s_3g_3+g_4^2)\\
\{g_3,f_4+f_5\} &=& g_1+g_2+2s_3g_3+2g_4^2\\
\{g_4,f_4+f_5\} &=& -2s_3g_4\,.
\end{eqnarray}
We see that $\{g_1+g_2-2g_3,f_4+f_5\}=0$, and if we trace back all the
dependencies on moments, we find that
\begin{equation} \label{Casimirg}
 g_1+g_2-2g_3=U_1
\end{equation}
is, in fact, the quadratic Casimir. The remaining independent variables can
conveniently be chosen as $g_1+g_2$, $g_1-g_2$ and $g_4$, with mutual Poisson
brackets
\begin{eqnarray}
 \{g_1+g_2,g_4\} &=& g_1-g_2\\
 \{g_1-g_2,g_4\} &=& g_1+g_2-2s_3g_3+2\frac{1+s_3}{1-s_3}g_4^2\\
 \{g_1+g_2,g_1-g_2\} &=& 4\frac{g_4}{1-s_3}(g_1+g_2+2s_3g_3+2 g_4^2)\,.
\end{eqnarray}

\paragraph{Final step:}
We now consider the flow $\partial/\partial s_3=\{\cdot,p_3\}$, using
(\ref{p3}):
\begin{eqnarray}
 \frac{\partial g_4}{\partial s_3}&=& \frac{g_4}{2(s_3-1)}\quad,\quad
 \frac{\partial(g_1-g_2)}{\partial s_3}=
 \frac{g_1-g_2}{2s_3(1-s_3)}\quad,\quad\nonumber\\
 \frac{\partial (g_1+g_2)}{\partial
   s_3}&=& \frac{g_1+g_2+2s_3g_3+2g_4^2}{2s_3(1-s_3)}= \frac{(g_1+g_2)(1+s_3)-
   s_3U_1+2g_4^2}{2s_3(1-s_3)}\,.
\end{eqnarray}
Solving these equations, we find that
\begin{eqnarray}
 h_1 &=& \frac{g_4}{\sqrt{s_3-1  }}\\
h_2 &=& (g_1-g_2)\sqrt{\frac{s_3-1}{s_3}}\\
h_3 &=& \frac{(1-s_3)(g_1+g_2)+s_3 U_1+2(1+s_3)(1-s_3)^{-1}g_4^2}{\sqrt{s_3}}\,,
\end{eqnarray}
in addition to $U_1$, are Poisson orthogonal to $s_3$ as well as $p_3$.  They
have closed brackets
\begin{equation}
 \{h_1,h_2\} = h_3\quad,\quad
\{h_1,h_3\} = -h_2\quad,\quad
\{h_2,h_3\} = 8h_1 U_1-32 h_1^3\,. \label{h2h3}
\end{equation}

As our final canonical momentum, we choose $p_4=h_1$. Its flow
equations
\begin{equation}
 \frac{\partial h_2}{\partial s_4}=-h_3\quad,\quad \frac{\partial
   h_3}{\partial s_4}=h_2
\end{equation}
have trigonometric solutions with a phase that can be set to zero by shifting
$s_4$. Therefore,
\begin{equation}
h_2=A(p_4) \cos(s_4)\quad,\quad
h_3=A(p_4) \sin(s_4)\,.
\end{equation}
The required Poisson brackets provide a condition on the function
$A(p_4)$,
\begin{equation}
A(p_4) \frac{{\rm d} A(p_4)}{{\rm d} p_4}=-8 p_4 U_1+32 p_4^3\,,
\end{equation}
solved by
\begin{equation} \label{Ap4}
A(p_4)=\sqrt{U_2-8p_4^2 U_1+16 p_4^4}\,.
\end{equation}
The new free parameter $U_2$ is a constant and is our second Casimir variable.

\paragraph{Casimir--Darboux variables:} Inverting all intermediate relations,
we obtain the moments in terms of Casimir--Darboux variables,
\begin{eqnarray}
\Delta(q_1^2)&=&s_1^2\quad,\quad
\Delta(q_1 \pi_1)= s_1 p_1\\
\Delta(\pi_1^2)&=&p_1^2+\frac{\Phi(s_3,p_3,s_4,p_4)}{s_1^2}
\end{eqnarray}
with
\begin{eqnarray}
\Phi(s_3,p_3,s_4,p_4)&=&-\frac{s_3+1}{s_3-1}p_4^2-4 s_3 \sqrt{s_3-1}p_3 p_4+4
s_3^2\left(s_3-1\right)p_3^2\\ 
&&+\frac{1}{2}\frac{s_3}{s_3-1} U_1\nonumber\\
&&-\frac{1}{2}\frac{\sqrt{s_3}}{s_3-1}\sqrt{U_2-8p_4^2 U_1+16
  p_4^4}\left(\sqrt{s_3-1}\cos{(s_4)}+\sin{(s_4)}\right)\,, \nonumber 
\end{eqnarray}
for moments of the second classical pair of degrees of freedom,
\begin{eqnarray}
\Delta(q_2^2)&=&s_2^2\quad,\quad
\Delta(q_2 \pi_2)= s_2 p_2\\
\Delta(\pi_2^2)&=&p_2^2+\frac{\Gamma(s_3,p_3,s_4,p_4)}{s_2^2}
\end{eqnarray}
with
\begin{eqnarray} \label{Gamma}
\Gamma(s_3,p_3,s_4,p_4)&=&-\frac{s_3+1}{s_3-1}p_4^2+4 s_3 \sqrt{s_3-1}p_3
p_4+4 s_3^2\left(s_3-1\right)p_3^2\\ 
&&+\frac{1}{2}\frac{s_3}{s_3-1} U_1\nonumber\\
&&-\frac{1}{2}\frac{\sqrt{s_3}}{s_3-1}\sqrt{U_2-8p_4^2 U_1+16
  p_4^4}\left(-\sqrt{s_3-1}\cos{(s_4)}+\sin{(s_4)}\right)\,, \nonumber
\end{eqnarray}
and
\begin{eqnarray}
\Delta(\pi_1 \pi_2)&=&\frac{p_1
  p_2}{\sqrt{s_3}}+\sqrt{\frac{s_3-1}{s_3}}
\left(\frac{p_2}{s_1}-\frac{p_1}{s_2}\right)p_4\\ 
&&-2
\sqrt{s_3}\left(s_3-1\right)
\left(\frac{p_1}{s_2}+\frac{p_2}{s_1}\right)p_3+\frac{\left(3
    s_3-1\right)}{s_1 s_2 \sqrt{s_3}\left(s_3-1\right)}p_4^2\nonumber\\ 
&&-4\frac{\left(s_3-1\right)s_3^{3/2}}{s_1 s_2}p_3^2-\frac{\sqrt{s_3}}{2 s_1
  s_2 \left(s_3-1\right)}U_1\nonumber\\ 
&&+\frac{s_3}{2 s_1 s_2 \left(s_3-1\right)}\sin{(s_4)}\sqrt{U_2-8p_4^2 U_1+16
  p_4^4}\nonumber\\ 
\Delta(\pi_1 q_2)&=&\frac{p_1
  s_2}{\sqrt{s_3}}+\sqrt{\frac{s_3-1}{s_3}}\frac{s_2}{s_1}p_4-2
\left(s_3-1\right)\sqrt{s_3}\frac{s_2}{s_1}p_3\\ 
\Delta(\pi_2 q_1)&=&\frac{p_2
  s_1}{\sqrt{s_3}}-\sqrt{\frac{s_3-1}{s_3}}\frac{s_1}{s_2}p_4-2
\left(s_3-1\right)\sqrt{s_3}\frac{s_1}{s_2}p_3\\ 
\Delta(q_1 q_2)&=& \frac{s_1 s_2}{\sqrt{s_3}}
\end{eqnarray}	
for the cross-covariances.

\paragraph{Canonical transformation:}
We can change our Darboux coordinates by canonical transformations. An
intersting example is suggested by the trigonometric form in which $s_4$
appears in the equations derived so far, which can be extended to $s_3$ by
using the canonical pair
\begin{equation}
 \beta= \arctan\sqrt{s_3-1} \quad,\quad p_{\beta}= 2s_3\sqrt{s_3-1}p_{3}\,.
\end{equation}
Computing $s_3=1+\tan^2\beta=1/\cos^2\beta$, we see that the new variable
$\beta$ interprets the cross-correlation
\begin{equation}
 \frac{\Delta(q_1q_2)}{\sqrt{\Delta(q_1^2)\Delta(q_2^2)}}=\frac{1}{\sqrt{s_3}}=
 \cos\beta
\end{equation}
as an angle.  Uncorrelated canonical pairs are therefore orthogonal to each
other in the sense that $\cos\beta=0$. 

Because $s_4$ already appears in trigonometric functions in our realization,
we rename it by defining
\begin{equation}
 \alpha=s_4\quad,\quad p_{\alpha}=p_4\,.
\end{equation}
The canonical mapping then takes the form
\begin{eqnarray} 
\Delta(q_1^2)&=&s_1^2\quad,\quad 
\Delta(q_1 \pi_1)= s_1 p_1\quad,\quad
\Delta(\pi_1^2)=p_1^2+\frac{\Phi}{s_1^2} \label{DeltaPhi}\\
\Delta(q_2^2)&=&s_2^2\quad,\quad
\Delta(q_2 \pi_2)= s_2 p_2\quad,\quad
\Delta(\pi_2^2)=p_2^2+\frac{\Gamma}{s_2^2}
\end{eqnarray}
where 
\begin{eqnarray}
\Phi(\beta,p_{\beta},\alpha,p_{\alpha})&=& (p_{\alpha}-p_{\beta})^2\label{Phi}\\
&&+\frac{1}{2
  \sin(\beta)^2}\left(U_1-4 p_{\alpha}^2-\sqrt{U_2-U_1^2+(U_1-4
    p_{\alpha}^2)^2}\sin(\alpha+\beta)\right) \nonumber\\
\Gamma(\beta,p_{\beta},\alpha,p_{\alpha})&=&(p_{\alpha}+p_{\beta})^2
\label{Gammabeta}\\  
&&+\frac{1}{2
  \sin(\beta)^2}\left(U_1-4 p_{\alpha}^2-\sqrt{U_2-U_1^2+(U_1-4
    p_{\alpha}^2)^2}\sin(\alpha-\beta)\right)\,,\nonumber
\end{eqnarray}
as well as
\begin{eqnarray}
\Delta(\pi_1 \pi_2)&=&p_2p_2 \cos(\beta)-\frac{\cos(\beta)}{s_1
  s_2}p_{\beta}^2+\frac{\cos(\beta)+2 \cot(\beta)\csc(\beta)}{s_1
  s_2}p_{\alpha}^2\\
&&-\sin(\beta)p_{\beta}\left(\frac{p_2}{s_1}+
\frac{p_1}{s_1}\right)+p_{\alpha}
\sin(\beta)\left(\frac{p_2}{s_1}-\frac{p_1}{s_2}\right)\nonumber\\
&&-\frac{\cot(\beta)\csc(\beta)}{s_1
  s_2}U_1+\frac{\csc(\beta)^2\sin(\alpha)}{2 s_1 s_2}\sqrt{16 p_{\alpha}^2-8
  p_{\alpha}^2 U_1+U_2}\nonumber\\ 
\Delta(\pi_1 q_2)&=&p_1 s_2
\cos(\beta)+\sin(\beta)\frac{s_2}{s_1}\left(p_{\alpha}-p_{\beta}\right)\\  
\Delta(\pi_2 q_1)&=&p_2 s_1
\cos(\beta)+\sin(\beta)\frac{s_1}{s_2}\left(p_{\beta}+p_{\alpha}\right)\\  
\Delta(q_1 q_2)&=&s_1 s_2 \cos(\beta) \, . \label{Deltaq1q2}
\end{eqnarray}	

\section{Applications}

As shown in the preceding section, the inclusion of moments in semiclassical
truncations leads to several new degrees of freedom. In this section, we
highlight some of the physical effects implied by them. At the same time, we
show that the form in which canonical variables appear in various realizations
of the moment algebras suggests truncations to smaller canonical subsystems
which are easier to analyze by analytic means and often show physical effects
more intuitively.

\subsection{Partition and two-point function of a free massive scalar field}
\label{s:thermo}

Our first example is an application of the second-order mapping (\ref{sps}),
rederived here from \cite{GaussianDyn,QHDTunneling}, to a free field theory.
We start with the Hamiltonian,
\begin{equation}
H=\int {\rm d}x \left(\frac{1}{2}\pi^2+\frac{1}{2}\left(\partial_x
    \phi\right)^2+\frac{1}{2}m^2 \phi^2\right)
\end{equation}
of a 1-dimensional real scalar field with mass $m$.  We transform to momentum
space by writing
\begin{equation}
\phi_k =\frac{1}{\sqrt{2\pi}}\int {\rm d}x \, \phi(x)e^{-i k x}\quad,\quad
\pi_k = \frac{1}{\sqrt{2\pi}}\int {\rm d}x \, \pi(x)e^{-i k x}
\end{equation}
with a real wave number $k$. Reality of $\phi(x)$ and $\pi(x)$ implies that
$\phi_k^*=\phi_{-k}$ and $\pi_k^*=\pi_{-k}$.

If we assume that the spatial manifold with
coordinate $x$ is compact and of length $2\pi$, thus describing a scalar field
on a unit circle, $k$ takes integer values and we have finite Poisson brackets
\begin{equation}
 \{\phi_k,\pi_{k'}\}= \delta_{kk'}
\end{equation}
replacing the field-theory Poisson brackets $\{\phi(x),\pi(y)\}=\delta(x-y)$
in the position representation. Each mode with fixed $k$ is then described by
an independent canonical pair $(\phi_k,\pi_k)$, which can easily be quantized
to a pair $(\hat{\phi}_k,\hat{\pi}_k)$ of operators.

The classical reality condition implies the adjointness relations
\begin{equation}
\hat{\phi}_k^{\dagger}=\hat{\phi}_{-k}\quad,\quad
\hat{\pi}^{\dagger}_{k}=\hat{\pi}_{-k}\,.
\end{equation}
The Hamilton operator can therefore be expressed as
\begin{equation}
\hat{H}=\frac{1}{2}\sum_{k=-\infty}^{\infty}\left(\hat{\pi}_k\hat{\pi}_{k}^{\dagger}
+\omega_k^2  \hat{\phi}_{k}\hat{\phi}_k^{\dagger}\right)
\end{equation}
with $\omega_k=\sqrt{m^2+k^2}$.
A further transformation,
\begin{equation}
\hat{\phi}_k =\frac{1}{2}\left(\hat{\phi}_k^{\rm R}-i \hat{\phi}_k^{\rm
    L}\right)\quad,\quad 
\hat{\pi}_k = \frac{1}{2}\left(\hat{\pi}_k^{\rm R}+i \hat{\pi}_k^{\rm L}\right)\,,
\end{equation}
explicitly decouples left and right-moving modes, $\hat{\phi}_k^{\rm L}$ and
$\hat{\phi}_k^{\rm R}$, respectively. 
The Hamilton operator then reads
\begin{equation}
\label{hammer}
\hat{H}=\frac{1}{2}\sum_{k=-\infty}^{\infty}\left(\left(\hat{\pi}^{\rm
      R}_{k}\right)^2+\left(\hat{\pi}^{\rm
      L}_{k}\right)^2+\frac{1}{4}\omega_k^2\left(\hat{\phi}^{\rm
      R}_{k}\right)^2+\frac{1}{4}\omega_k^{2}\left(\hat{\phi}^{\rm
    L}_{k}\right)^{2}\right)\,.
\end{equation}

\subsubsection{Partition function}

Since all the modes decouple and have harmonic Hamiltonians, the mapping for a
single degree of freedom at the second order provides an exact effective
description in any state in which cross-correlations between different modes
vanish.  In the absence of interaction terms in the Hamiltonian, the latter
condition is satisfied in the ground state. More generally, we can also
consider ensemble averages in finite-temperature states. Since
cross-correlations do not contribute the the energy of our non-interacting
system, they will not be affected by a turning on a finite temperature.
Moreover, correlations in harmonic systems have oscillatory solutions around
zero and therefore vanish in an ensemble average. 

Mode fluctuations parameterized by the canonical variable $s_k$ with momentum
$p_k$ and Casimir $U_k$, by contrast, are bounded from below by the
uncertainty relation and do not average to zero.  For every fixed mode and at
finite temperature $T$, we can compute the partition function
\begin{equation}
  \mathcal{Z}(\beta,\omega_k,\lambda)=
  \int_{0}^{\infty}\int_{-\infty}^{\infty}\int_{U_{\rm min}}^{\infty}\, 
  {\rm d}s_k \,{\rm d}p_k \, {\rm d}U_k 
 \exp{\left(-\beta \left(\frac{1}{2}p_k^2+\lambda
        \frac{U_k}{2 
          s_k^2}+\frac{1}{8}\omega_k^2 s_k^2\right)\right)}\,, 
\end{equation}
where $\beta=1/k_{\rm B}T$ and $U_{\rm min}=\hbar^2/4$ and we have restricted
$s_k$ to positive values.  We have inserted the auxiliary parameter $\lambda$ in
anticipation of an application below in which a $\lambda$-derivative of ${\cal
  Z}$ will give us the ensemble average of the quantum uncertainty $U_k$. For
all other purposes, we use the physical value $\lambda=1$. If we perform the
$U_k$-integral before the $s_k$-integral, the partition function
\begin{equation}
\mathcal{Z}(\beta,\omega_k,\lambda)=4\pi \lambda^{-1}
\omega_k^{-3}\beta^{-3}\left(2+\beta \omega_k 
  \sqrt{U_{\rm min} \lambda }\right)\exp{\left(-\frac{1}{2}\beta \omega_k
    \sqrt{U_{\rm min}\lambda }\right)} 
\end{equation}
can be obtained in closed form.

A derivative by $\omega_k$ (at $\lambda=1$) results in the
ensemble averages
\begin{equation} \label{sE}
\langle (s_k^{\rm R})^2\rangle_{\rm E}=\langle (s_k^{\rm L})^2 \rangle_{\rm
  E}=\frac{12}{\omega_k^2 
  \beta}+\frac{U_{\rm min}\beta}{1+\frac{1}{2}\sqrt{U_{\rm min}}\omega_k \beta}.
\end{equation}
of dispersions in a thermal state.  Moreover, the average energy per mode is
\begin{equation}
\langle E_k \rangle_{\rm E}=-\frac{\partial \log{\mathcal{Z}}}{\partial
  \beta}=\frac{12 + \beta \omega_k \left(6 \sqrt{U_{\rm min}}+U_{\rm min}\omega_k
    \beta\right)}{2 \beta \left(2+\beta \sqrt{U_{\rm min}}\omega_k\right)}.
\end{equation}
In the limit $T\rightarrow 0$, the value
\begin{equation}
\langle E_k \rangle_{E}=\sqrt{U_{\rm min}}\frac{\omega_k}{2}
\end{equation}
agrees with the ground-state energy if we use $U_{\rm min}=\hbar^2/4$, noting
that a single mode used here appears with frequency $\omega_k/2$ in
(\ref{hammer}). (The combination of $\phi_k^{\rm R}$ and $\phi_k^{\rm L}$ has
the standard harmonic-oscillator energy $\frac{1}{2}\hbar\omega_k$ on
average.) Finally, the ensemble average of the quantum uncertainty in a
thermal state can be determined as
\begin{equation}
\langle U_k \rangle_{E}=\frac{8}{\beta^2
  \omega_k}\frac{1}{\mathcal{Z}}\left.\frac{\partial^2 \mathcal{Z}}{\partial
  \omega_k \partial \lambda}\right|_{\lambda=1}=U_{\rm min}+\frac{24}{\beta^2
  \omega_k^2}+\frac{4 U_{\rm min}}{2+\sqrt{U_{\rm min}}\beta \omega_k} \,,
\end{equation}
which approaches $U_{\rm min}$ as $T\rightarrow 0$. For $T\not=0$, $\langle
U_k\rangle_{\rm E}>U_{\rm min}$ in a mixed, finite-temperature state. The
difference $U-U_{\rm min}$ is therefore an impurity parameter in this
situation, which is in agreement with our discussion in Sec.~\ref{s:Purity}
and the fact that the Casimir $U$ only appears in the second-order moment
$\Delta(\pi^2)$. 

We see that canonical variables for semiclassical truncations can give easy
access to thermodynamical quantities by rewriting a quantum statistical system
in the form of a classical system. The canonical nature of variables
parameterizing quantum moments makes it possible to determine the correct
phase-space volume for the partition function.

\subsubsection{Two-point function}

We extend the definition of moments to our field theory by applying the
quantum-mechanics definition to each mode $\phi_k$. Introducing
$\widehat{\delta\phi}_k= \hat{\phi}_k-\langle\hat{\phi}_k\rangle_{\rm Q}$, we
then have $\Delta(\phi_k\phi_{k'})=
\langle\widehat{\delta\phi}_k\widehat{\delta\phi}_{k'}\rangle_{\rm Q}$, from
which we can obtain correlations in the position representation by Fourier
transformation. In these definitions, we have explicitly indicated that
expectation values $\langle\cdot\rangle_{\rm Q}$ refer to a quantum state as
opposed to the ensemble average used in (\ref{sE}).

The two-point function
\begin{eqnarray*}
\langle \Delta\left(\phi(x)\phi(y)\right)\rangle_{\rm E} &=& \sum_{k k'}\langle
\langle \widehat{\delta\phi}_k \widehat{\delta\phi}_{k'} \rangle_{\rm
  Q}\rangle_{\rm   E} e^{i k  x}e^{i k'  y}\\ 
 \\ 
&=&\frac{1}{4}\sum_{k k'}\left\langle
  \left\langle\left(\widehat{\delta\phi}{}_{k}^{\rm R}-i 
  \widehat{\delta\phi}{}^{\rm L}_{k}\right)
\left(\widehat{\delta\phi}{}_{k'}^{\rm R}-i 
  \widehat{\delta\phi}{}^{\rm L}_{k'}\right)\right\rangle_{\rm Q}\right\rangle_{\rm
E} e^{ikx}e^{ik'y}
\end{eqnarray*}
combines both types of averages. We can simplify the double summation using
$\widehat{\delta\phi}{}^{\rm L}_{-k}=-\widehat{\delta\phi}{}^{\rm L}_{k}$, which
follows from the adjointness relation for $\hat{\phi}_k$.  Using zero
cross-covariances between the modes as well as the fact that the fluctuations
only depend on the wave number $k$ but not on whether the mode is left or
right-moving, the double summation is then reduced to
\begin{equation}
\langle
\Delta\left(\phi(x)\phi(y)\right)\rangle_{\rm E}=\frac{1}{2}\sum_k\langle\langle
\widehat{\delta\phi}{}^{\rm R}_{k}\widehat{\delta\phi}{}^{\rm R}_{k}\rangle_{\rm
  Q}\rangle_{\rm E} \cos{\left(k(x-y)\right)} \,.
\end{equation}
Inserting (\ref{sE}), we obtain
\begin{equation}
\langle
\Delta\left(\phi(x)\phi(y)\right)\rangle_{\rm E}=
\frac{1}{2}\sum_k\left(\frac{12}{\omega_k^2 
    \beta}+\frac{U_{\rm min}\beta}{1+\frac{1}{2}\sqrt{U_{\rm min}}\omega_k
    \beta}\right) \cos{\left(k(x-y)\right)}\,. 
\end{equation}

In the limit in which the radius of the circle goes to infinity, we can
replace $\sum_k$ by $(2\pi)^{-1}\int {\rm d} k$, such that
\begin{equation}
\langle \Delta\left(\phi(x)\phi(y)\right)\rangle_{\rm E}=
\frac{1}{2}\int \frac{{\rm
    d}k}{2 \pi}\left(\frac{12}{\omega_k^2 \beta}+\frac{U_{\rm
      min}\beta}{1+\frac{1}{2}\sqrt{U_{\rm min}}\omega_k \beta}\right)
\cos{\left(k(x-y)\right)}\, .
\end{equation}
It is instructive to consider the low-temperature limit $\beta \rightarrow
\infty$. The result,
\begin{eqnarray}
\lim_{\beta\to\infty}\langle \Delta\left(\phi(x)\phi(y)\right)\rangle_{\rm
  E}&=&\hbar\int \frac{{\rm d}k}{4 \pi \omega_k}\cos{(k(x-y))}\nonumber\\
&=& \frac{\hbar}{2\pi} K_0(m|x-y|)
\end{eqnarray}
with a Bessel function $K_0$, agrees exactly with the equal-time two-point
function obtained using path integral methods.

We can also consider the case where the temperature is nonzero but still small
enough for the semi-classical approximation to be valid. Taylor expanding the
integrand about $\beta=\infty$, the first-order temperature correction to the
two-point function is $8/(\omega_k^2 \beta)$:
\begin{equation}
\langle \Delta\left(\phi(x)\phi(y)\right)\rangle_{\rm E}=\frac{\hbar}{2\pi}
K_0(m|x-y|)
+\frac{9 k T}{4 m}\exp{(-m|x-y|)}+O(T^2)\,.
\end{equation}
The asymptotic behavior $K_0(z)\sim \sqrt{\frac{1}{2}\pi/z}\; e^{-z}$ for large
$z$ shows that the term linear in the temperature decreases more slowly with
the distance than the temperature-independent term. For large-distance
correlations, this correction from a non-zero temperature may therefore be
relevant.

\subsection{Closure conditions}

Our third and fourth order mappings suggest new closure conditions (in the
sense of \cite{Closure}) that can be used to describe moments by a small
number of parameters. In particular, we may assume that the second-order
fluctuation parameter $s$ contributes to higher-order moments such that
$\Delta(q^n)=s^n$, at least for even $n$. For the third-order moments
$\Delta(q^3)$ in the fourth-order truncation, we have seen that the cubic
dependence on $s_i$ is multiplied by a free parameter, given by the Casimir
variable $C$, which is lacking in even-order moments $\Delta(q^2)$ and
$\Delta(q^4)$. Since odd-order moments are often sub-dominant, for instance in
the family of Gaussian states, we can set $C=0$ and assume that this
behavior extends to higher orders. These considerations suggest the closure
conditions
\begin{equation}
 \Delta(q^n)=\left\{\begin{array}{cl}s^n & \mbox{for even }n\\ 0 & \mbox{for
       odd }n\end{array}\right.
\end{equation}
for all moments, replacing a truncation to finite order. In an effective
Hamiltonian, we then obtain the {\em all-orders effective potential}
\begin{equation} \label{AllOrders}
 V_{\rm all-orders}(q,s)= V(q)+ \frac{U}{2ms^2} +\sum_n \frac{1}{(2n)!} \frac{{\rm
     d}^{2n}V(q)}{{\rm d}q^{2n}} s^{2n}=
 \frac{U}{2ms^2}+\frac{1}{2}\left(V(q+s)+V(q-s)\right) 
\end{equation}
for a classical potential $V(q)$. The Casimir variable $U$ may be set equal to
the minimum value $\hbar^2/4$ allowed by the uncertainty relation.

\subsubsection{Non-differentiable potentials}

Semiclassical physics is usually based on an expansion which requires a smooth
potential. Our all-orders effective potential, by contrast, explicitly sums up
a perturbative series and expresses quantum effects via finite shifts of the
classical potential. It can therefore be applied to potentials that are not
smooth or not even differentiable.

As an example, consider the potential $V(q)=|q|$. In particular, we can check
the ground state energy. In the static case of zero momentum (and using atomic
units in which $\hbar=1$ and $m=1$), we have
\begin{equation}
V_{\rm all-orders}(q,s)=\frac{1}{8 s^2}+\frac{1}{2}\left(|q+s|+|q-s|\right)\,.
\end{equation}
This function has a minimum at $q=0$ and $s=2^{-2/3}$, and the minimum value
is $E_{\rm ground}=0.94$. We can calculate the exact value of the ground state
energy using a truncated oscillator basis. The result is $E_{\rm ground}^{\rm
  exact}=0.81$.

It is possible obtain this non-differentiable potential as a limit of a
differentiable one.  To this end, consider the Hamiltonian
\begin{equation}
H=\sqrt{1+\pi^2}+\frac{1}{2} q^2
\end{equation}
which can be interpreted as describing a relativistic particle with
position-dependent mass $\sqrt{H^2-\pi^2}=\sqrt{1+\frac{1}{2}q^2}$.  After a
simple canonical transformation $(q,\pi)\mapsto (-\pi,q)$ the Hamiltonian
\begin{equation}
H=\frac{\pi^2}{2 }+\sqrt{1+q^2}
\end{equation}
appears in standard form for a non-relativistic system.  Now the
all-orders effective potential with $U=1/4$ is given by
\begin{equation}
  V_{\rm all-orders}(q,s)=\frac{1}{8
    s^2}+\frac{1}{2}\left(\sqrt{1+(q+s)^2}+\sqrt{1+(q-s)^2}\right) 
\end{equation}
and minimized when $q=0$. Minimizing
\begin{equation}
V_{\rm all-orders}(0,s)=\frac{1}{8  s^2}+\sqrt{1+s^2}
\end{equation}
with respect to $s$, we find the minimum value
\begin{equation}
E_{\rm ground}=1.47\,.
\end{equation}
The exact ground state energy is
\begin{equation}
E^{\rm exact}_{\rm ground}=1.44.
\end{equation}
The agreement here is better than in the preceding example, which can be
interpreted as a limit of a potential in which $\sqrt{1+q^2}$ is replaced by
$\lim_{d\to 0}\sqrt{d+q^2}$.

\subsubsection{Canonical tunneling in polynomial potentials}

The regimes of validity of the all-orders potential can be tested in the case
of tunneling escape. For this purpose, we consider a fourth-order polynomial
potential in order to describe tunneling escape from a metastable state:
\begin{equation}
\label{eq:polynomial-potential}
V_{\rm poly}(q) =\frac{27}{4}V_{\rm top} \gamma q^2
\left(q-1\right)\left(q-\frac{1}{\gamma}\right)\, , 
\end{equation}
where $V_{\rm top}$ is a parameter that controls the height of the barrier and
$\gamma$ controls the location of the global minimum of this potential. When
$\gamma$ is small, this potential has the following approximate critical
points with the corresponding potential values: The top of the barrier is
characterized by
\begin{equation}
q_{\rm top}\approx\frac{2}{3} \quad,\quad V_{\rm poly}(q_{\rm top})
=V_{\rm top}
\end{equation}
and the global minimum is characterized by
\begin{equation}
q_{\rm min}\approx\frac{3}{4\gamma}  \quad,\quad V_{\rm poly}(q_{\rm
  min}) \approx -\frac{729\, V_{\rm top}}{1024\gamma^3}\,.
\end{equation}
In addition to the global minimum, there is a local minimum at $q=0$
with $V_{\rm poly}(0) = 0$.

Classically, if the particle starts close to the local minimum at $q=0$ with
an energy less than $V_{\rm top}$, the particle will remain confined. However if
quantum degrees of freedom are taken into account, we know that the particle
can tunnel through the barrier and into the lower basin. We can account for
this modified dynamics using second-order variables if the barrier is
sufficiently small. If the barrier is large, higher-order corrections need
to be taken into account in order to see tunneling. The all-orders effective
potential, given by
\begin{equation}
\label{eq:all-orders-polynomial}
V_{\rm all-orders}(q,s) = \frac{U}{2ms^2} +\frac{1}{2}\left( V_{\rm poly}(q+s) +
  V_{\rm poly}(q-s)\right)
\end{equation}
includes some of the terms that result from higher-order moments.

For escape from a metastable state, the particle is initially at the local
minimum at $q=0$, around which
\begin{equation}
V_{\rm poly}(q)\approx\frac{27}{4}V_{\rm top}q^2\,.
\end{equation}
For this quadratic approximation, the effective potential is
\begin{equation}
V_{\rm eff}(q,s) \approx \frac{27}{4}V_{\rm
  top}\left(q^2+s^2\right)+\frac{U}{2 s^2} \, .
\end{equation}
This potential has a minimum at
\begin{equation}\label{incondions}
q=0\quad,\quad
s=\left(\frac{2 U}{27 V_{\rm top}}\right)^{1/4}
\end{equation}
which give the approximate ground state energy
\begin{equation}
V_{0}\approx\frac{3}{8}\sqrt{\frac{3 U}{V_{\rm top}}}\left(V_{\rm top}+2\right)\,.
\end{equation}

\begin{figure}[htbp]
\begin{center}
\includegraphics[scale=0.6]{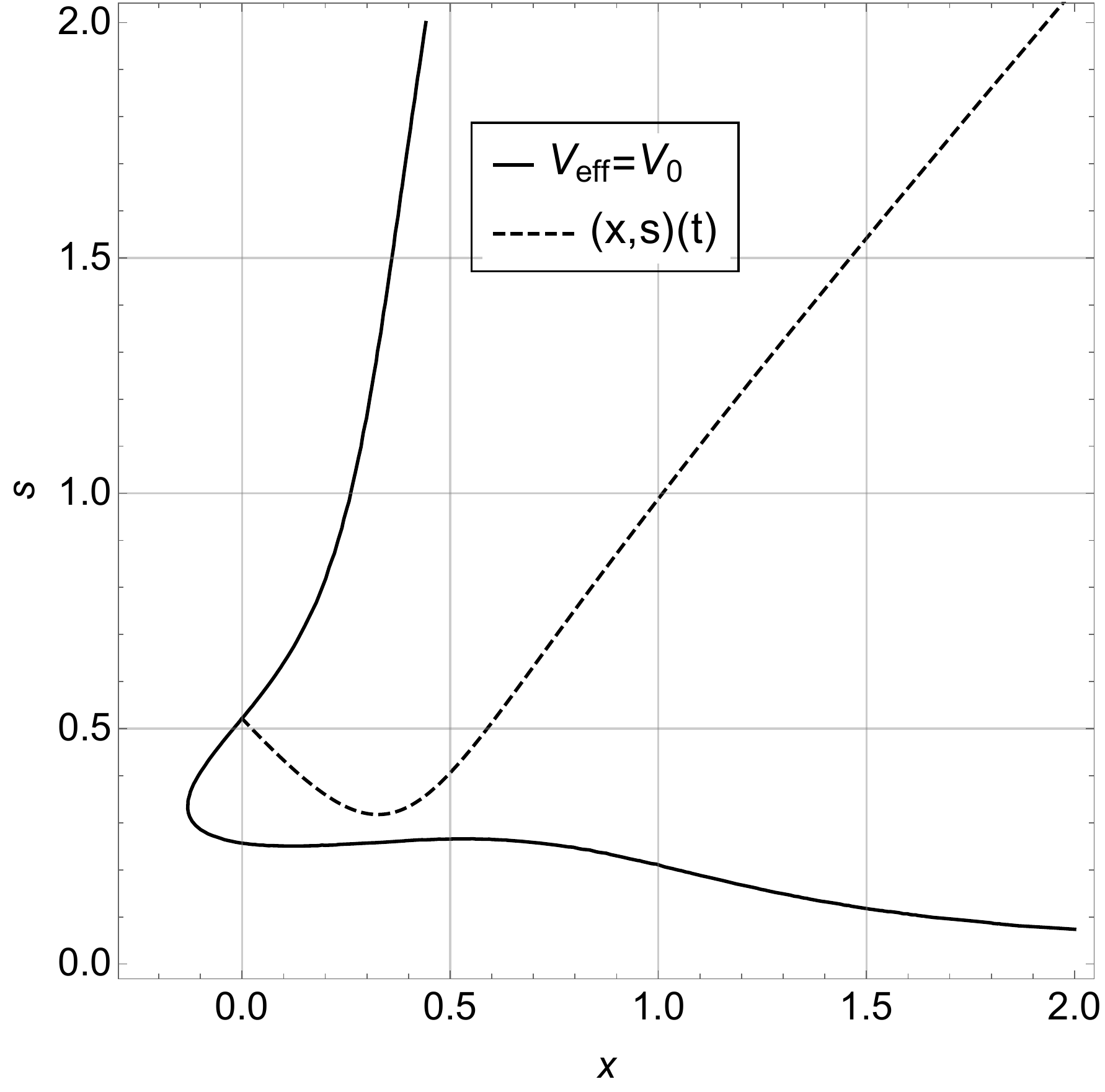}
\caption{Dynamics in the all-orders effective potential
  (\ref{eq:all-orders-polynomial}): The potential is represented by its
  ground-state equipotential curve $V_{\rm{eff}}=V_{0}$ (solid line), together
  with a tunneling trajectory starting from the local minimum (dashed
  line). For this plot we chose the parameters $V_{\rm top}=1$, $\gamma=0.1$,
  $U=1/4$. The ``extra dimension'' given by the fluctuation parameter $s$
  provides  the particle with an escape route around the classical barrier,
  without violating energy conservation.}
\label{fig:Contourplot}
\end{center}
\end{figure}

Given the initial conditions (\ref{incondions}) we can track the particle
dynamics numerically; see Fig.~\ref{fig:Contourplot}. If the parameter $V_{\rm
  top}$ becomes large the particles no longer tunnels if one only considers
the second-order canonical mapping. Second-order dynamics can provide good
approximations in certain regimes, but for deep tunneling we need an extension
to higher orders. The all-orders effective potential is then useful for
understanding the escape from a local minimum in deep tunneling situations.

\begin{figure}[htbp]
\begin{center}
 \includegraphics[scale=0.6]{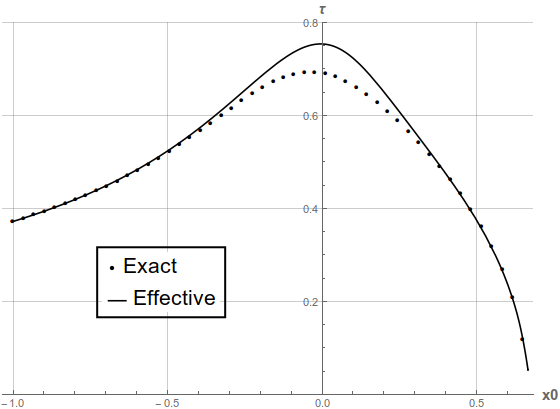}
 \caption{Tunneling times as a function of the starting position, for an exact
   calculation and the all-orders potential, respectively. 
   There is good agreement, with larger discrepancies close to the origin
   where we have deep tunneling.}
\label{fig:tunneling}
\end{center}
\end{figure}

\begin{figure}[htbp]
\begin{center}
\includegraphics[scale=0.6]{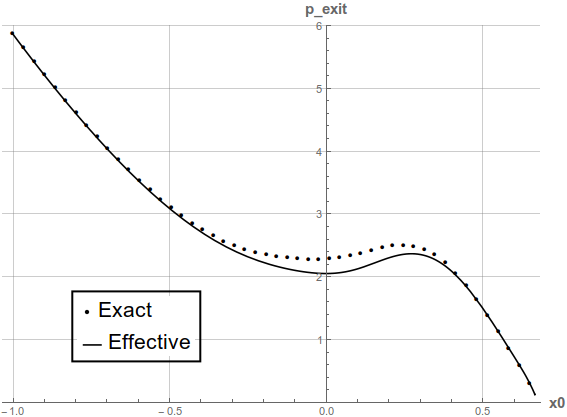}
\caption{The exit momentum of the particle as a function of the initial
  position.}
\label{fig:momentum}
\end{center}
\end{figure}

Using the all-orders potential, we estimate the tunneling time as a function
of the tunnel exit position of the particle, which corresponds to the particle
position around the critical point $q_{\rm{top}} \approx 2/3$.
Figures~\ref{fig:tunneling} and \ref{fig:momentum} show numerical comparison
of the canonical tunneling time and the exit momentum of the particle, using
the all-orders potential and exact solutions, respectively.

\begin{figure}[htbp]
\begin{center}
\includegraphics[scale=0.6]{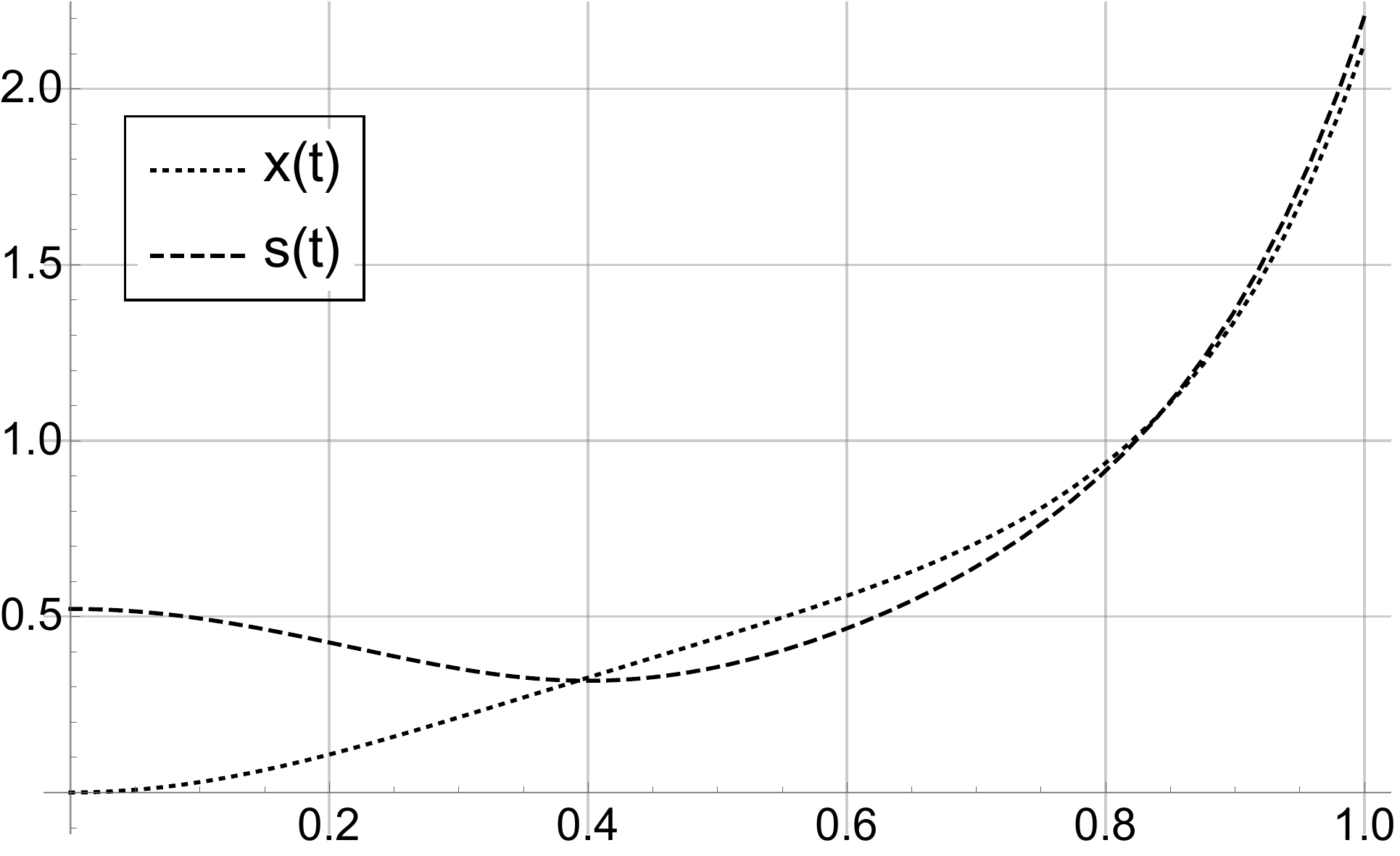}
\caption{Trajectories of the tunneling coordinate $q$ and its fluctuation $s$
  for the all-orders effective potential \eqref{eq:all-orders-polynomial}.}  
\label{fig:position-fluc}
\end{center}
\end{figure}

In \cite{Ionization} we used the all-orders effective potential for atomic
systems, based on the all-orders closure condition. In a further approximation,
it was possible to eliminate some of the basic variables such that $s \approx
q$ inside the barrier. For the polynomial potential we can test the same
behavior by computing the evolution of the expectation value $q$ and its
fluctuation $s$.  As shown Fig.~\ref{fig:position-fluc}, the approximate
relationship between $q$ and $s$ during tunneling is maintained also here.

\begin{figure}[htbp]
\begin{center}
 \includegraphics[scale=0.6]{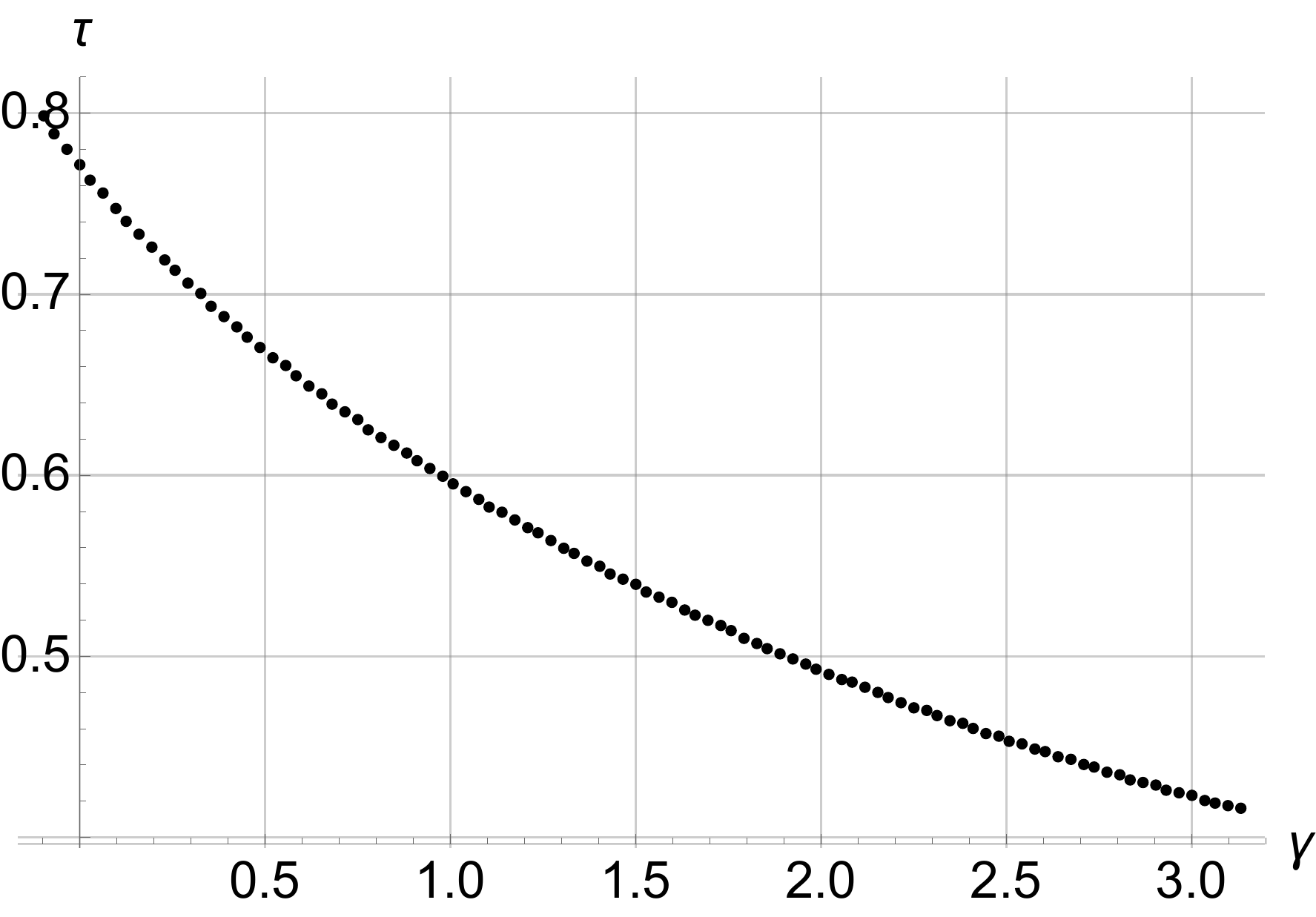}
 \caption{The tunneling time as a function of $\gamma$ in the
   potential \eqref{eq:polynomial-potential}.}
\label{fig:tunneling-gamma}
\end{center}
\end{figure}

Finally, it is interesting to note that the tunneling time can be sensitive to
the parameter $\gamma$ which specifies the location of the global minimum of
the classical potential \eqref{eq:polynomial-potential}. We
estimate the tunneling time in terms of $\gamma$, starting with
$\gamma=0.1$, as shown in Fig.~\ref{fig:tunneling-gamma}.

\subsection{Effective potentials}

Casimir--Darboux coordinates for moments, in combination with the effective
Hamiltonian (\ref{Heff}), allow us to identify the dynamics of a semiclassical
truncation with a dynamical canonical system. The classical momentum $\pi$
(derived from the momentum expectation value) is then accompanied by one or
more new momenta that parameterize fluctuations, correlations, and higher
moments. 

For a single classical pair of degrees of freedom to second semiclassical
order, the moments are quadratic in the new momentum $p_s$ with constant
coefficients. A dynamical system with standard kinetic term is therefore
obtained \cite{QHDTunneling}:
\begin{eqnarray}
 \langle\hat{H}\rangle &=& \frac{\langle\hat{\pi}^2\rangle}{2m}+V(\hat{q})=
 \frac{\pi^2+\Delta(\pi^2)}{2m}+ V(q)+ \frac{1}{2}V''(q) \Delta(q^2)+\cdots
 \nonumber \\
 &=&
 \frac{\pi^2}{2m}+ \frac{p_s^2}{2m}+ \frac{U}{2ms^2}+
 V(q)+\frac{1}{2}V''(q)s^2+\cdots
\end{eqnarray}
with effective potential
\begin{equation}
 V_{\rm eff}(q,s)= \frac{U}{2ms^2}+
 V(q)+\frac{1}{2}V''(q)s^2\,.
\end{equation}
Our third-order moments provide an extension to the next order, now with three
non-classical momenta. The first version, (\ref{Deltapi1}), is quadratic in
momenta but with coefficients depending on the configuration variables
$s_i$. The second version, (\ref{ansatz}), results in a simplified system with
constant coefficients in the extended kinetic term.

However, for two pairs of degrees of freedom, it is not possible to have
momentum fluctuations which are quadratic in Darboux momenta with constant
coefficients \cite{Bosonize}. The resulting effective theories are therefore
more involved in such cases. Nevertheless, it is possible to extract an
effective potential.  Using the Taylor expansion (\ref{Heff}) of the effective
Hamiltonian $\langle\hat{H}\rangle$ and setting all canonical momenta equal to
zero, we obtain an expression depending only on the canonical coordinates. We
do not require that the momenta vanish for all solutions of interest, which
would then be adiabatic, but rather extract a term from the effective
Hamiltonian that serves as an effective potential. For this purpose,
canonical variables are required in order to know which functions of the
moments should be considered momenta.

For two classical degrees of freedom to second semiclassical order, this
procedure leads to the effective potential
\begin{eqnarray} \label{Veff2}
  V^{(1)}_{\mathrm{eff}}(q_1,q_2,s_1,s_2,\alpha,\beta,U_1,U_2)&=&V(q_1,q_2)\\
&&+
 \frac{1}{4\sin^2(\beta)}
 \left(\frac{U_1-\sqrt{U_2}\sin(\alpha+\beta)}{s_1^2}+
  \frac{U_1-\sqrt{U_2}\sin(\alpha-\beta)}{s_2^2}\right)\nonumber\\
&&+\frac{1}{2}V_{11}(q_1,q_2) s_1^2+
 V_{12}(q_1,q_2) s_1 s_2
  \cos{\left(\beta\right)}+\frac{1}{2}V_{22}(q_1,q_2) s_2^2 \nonumber
\end{eqnarray}
We have used the notation $V_{ij}= \partial^2V/\partial q_i\partial q_j$, and
$V(q_1,q_2)$ is the classical potential. The two Casimir coordinates $U_1$ and
$U_2$ are constants of motion for any classical dynamics and can be considered
(state-dependent) parameters of the effective potential, while the remainder
in the effective Hamiltonian is a non-standard kinetic term.

We define the low-energy effective potential $V_{\rm low}(q_1,q_2)$
as the effective potential $V_{\rm eff}$ restricted to values of the moments
(that is, $s_1$, $s_2$, $\alpha$, $\beta$, $U_1$ and $U_2$) obtained in the
ground state of the interaction system.  We therefore determine the moments by
minimizing the effective potential with respect to $s_1$, $s_2$, $\alpha$,
$\beta$ and the two Casimir coordinates while keeping the classical-type
variables $q_1$ and $q_2$ free. 

In this process, we have to respect the boundaries imposed by uncertainty
relations. Since $W$ is linear in $U_1$ and $\sqrt{U_2}$, minimization sends
these two values to the boundary. (From (\ref{Ap4}), we know that $U_2>0$ for
$p_4=0$ to be possible.) The relevant boundary components, at zero momenta,
can be obtained from Heisenberg's uncertainty relation applied to each
canonical pair:
\begin{eqnarray} \label{relations}
\Phi(\beta,0,\alpha,0)&=&\frac{1}{2
  \sin{(\beta)}^2}\left(U_1-\sqrt{U_2}\sin{(\alpha+\beta)}\right)\geq
\frac{\hbar^2}{4}\\ 
\Gamma(\beta,0,\alpha,0)&=&\frac{1}{2
  \sin{(\beta)}^2}\left(U_1-\sqrt{U_2}\sin{(\alpha-\beta)}\right)\geq
\frac{\hbar^2}{4}\,.  \label{relation2}
\end{eqnarray}
For fixed $U_1$ and $U_2$, these two relations must be true for all $\alpha$
and $\beta$. Moreover, for any choice of $U_1$ and $U_2$ there must be
solutions of $\alpha$ and $\beta$ such that both relations are saturated: If
the coupling between the two degrees of freedom is turned off adiabatically we
expect saturation in the ground state. Since $U_1$ and $U_2$ are
constants of motion for any Hamiltonian, their values do not change during
this adiabatic decoupling. Therefore, any choice of $U_1$ and $U_2$ must allow
some solutions of $\alpha$ and $\beta$ such that the uncertainty relations are
saturated.

At saturation, we can subtract (\ref{relations}) and (\ref{relation2}) and
obtain
\begin{equation} \label{U2alpha}
 -\frac{1}{2}\sqrt{U_2} \frac{\cos(\alpha)}{\sin(\beta)}=0\,,
\end{equation}
and thus $U_2=0$ or $\cos(\alpha)=0$. In the latter case, the $U_2$-dependent
term in the effective potential,
\begin{equation}
 V_{U_2}= -\frac{\sqrt{U_2}\cos(\beta)}{4\sin^2(\beta)}
 \left(\frac{1}{s_1^2}+\frac{1}{s_2^2}\right)\,,
\end{equation}
is, for any classical potential, unbounded from below in $\sqrt{U_2}$ for any
$\beta$ such that $\cos(\beta)>0$. This solution of (\ref{U2alpha}) is
therefore ruled out by the condition that a stable ground state must exist for
a large class of classical potentials. We conclude that $U_2=0$. 

Given this solution, the smallest value of $U_1$ for which (\ref{relations})
can be fulfilled is $U_1=\hbar^2/2$. Therefore,
\begin{equation}
\Phi|_{p_3=p_4=U_2=0,U_1=\hbar^2/2}=\frac{\hbar^2}{4 \sin{\beta}^2} =
\Gamma_{p_3=p_4=U_2=0, U_1=\hbar^2/2} 
\end{equation}
from (\ref{Phi}) and (\ref{Gammabeta}).  The effective potential then reads
\begin{eqnarray}
V^{(2)}_{\mathrm{eff}}(q_1,q_2,s_1,s_2,\beta)&=&V(q_1,q_2)\nonumber\\
&&+\frac{\hbar^2}{8
  \sin{(\beta)}^2s_1^2}+\frac{\hbar^2}{8
  \sin{(\beta)}^2 s_2^2}+\frac{1}{2}V_{11} s_1^2\nonumber\\
&&+V_{12} s_1 s_2
\cos{\left(\beta\right)}+\frac{1}{2}V_{22} s_2^2 \,.
\end{eqnarray}
Although we have not minimized the potential in the direction of $\alpha$, the
$\alpha$-dependence has disappeared. There should, however, be a unique pure
state that corresponds to the ground state where the effective potential has
its minimum. Since minimization does not determine $\alpha$, it must be the
pure-state condition that fixes its value. This conclusion is in agreement
with our earlier discussion of impurity parameters: In the mapping
(\ref{DeltaPhi})--(\ref{Deltaq1q2}), $\alpha$ appears only in moments of the
form $\Delta(\pi_i\pi_j)$ which are not required to reconstruct a pure state
in the position representation.

Minimization by $s_1$, $s_2$ and $\beta$ gives us three equations:
\begin{eqnarray}
 0&=& \frac{\partial V^{(2)}_{\rm eff}}{\partial s_1} =
 -\frac{\hbar^2}{4 s_1^3\sin^2\beta}+ V_{11}s_1+ V_{12}s_2\cos\beta \label{Vs1}\\
 0&=& \frac{\partial V^{(2)}_{\rm eff}}{\partial s_2} =
 -\frac{\hbar^2}{4 s_2^3\sin^2\beta}+ V_{22}s_2+
 V_{12}s_1\cos\beta\label{Vs2}\\
0&=&\frac{\partial V^{(2)}_{\rm eff}}{\partial\beta}=
-\frac{\hbar^2(s_1^2+s_2^2)\cos\beta}{4 s_1^2s_2^2\sin^2\beta}-
V_{12}s_1s_2\sin\beta\,. \label{Vbeta}
\end{eqnarray}
Subtracting $s_2$ times (\ref{Vs2}) from $s_1$ times (\ref{Vs1}), we obtain
\begin{equation}\label{sin2}
 \sin^2\beta= \frac{\hbar^2}{4 s_1^2s_2^2}
 \frac{s_2^2-s_1^2}{V_{11}s_1^2-V_{22}s_2^2}\,.
\end{equation}
Using the sum of $s_2$ times (\ref{Vs2}) and $s_1$ times
(\ref{Vs1}), we derive
\begin{eqnarray} \label{cos1}
 4V_{12}^2s_1^2s_2^2\cos^2\beta&=&
 \left(\frac{\hbar^2(s_1^2+s_2^2)}{4 s_1^2s_2^2\sin^2\beta}-
   (V_{11}s_1^2+V_{22}s_2^2)\right)^2\\
&=& \left(\frac{s_1^2+s_2^2}{s_1^2-s_2^2} (V_{11}s_1^2-V_{22}s_2^2)+
  (V_{11}s_1^2+V_{22}s_2^2)\right)^2\\
&=& 4\frac{(V_{11}s_1^4-V_{22}s_2^4)^2}{(s_1^2-s_2^2)^2}\,.
\end{eqnarray}

Alternatively, we can derive $4V_{12}^2s_1^2s_2^2\cos^2\beta$ as follows: The
sum of $s_1\sin^2\beta$ times (\ref{Vs1}) and $\cos\beta\sin\beta$ times
(\ref{Vbeta}) implies
\begin{eqnarray} \label{Vbeta2}
 0&=& -\frac{\hbar^2(s_1^2+s_2^2)}{4 s_1^2s_2^2\sin^2\beta}+
 \frac{\hbar^2}{4 s_2^2}+ V_{11}s_1^2\sin^2\beta\\
&=& \frac{s_1^2+s_2^2}{s_1^2-s_2^2} (V_{11}s_1^2-V_{22}s_2^2)+
\frac{\hbar^2}{4} \frac{V_{11}-V_{22}}{V_{11}s_1^2-V_{22}s_2^2}
\end{eqnarray}
using (\ref{sin2}). This equation together with (\ref{sin2}) also gives us
\begin{eqnarray} \label{cos2}
4V_{12}^2s_1^2s_2^2 \cos^2\beta &=&
4V_{12}^2s_1^2s_2^2\left(1-\frac{2 \hbar^2}{s_1^2s_2^2}
  \frac{s_2^2-s_1^2}{V_{11}s_1^2-V_{22}s_2^2}\right)\\
&=& 4V_{12}^2\left(s_1^2s_2^2+
  (s_1^2+s_2^2)\frac{V_{11}s_1^2-V_{22}s_2^2}{V_{11}-V_{22}}\right)\\
&=& 4V_{12}^2\frac{V_{11}s_1^4-V_{22}s_2^4}{V_{22}-V_{11}}\,.
\end{eqnarray}
Equating (\ref{cos1}) and (\ref{cos2}), we have
\begin{equation}
 V_{11}s_1^4-V_{22}s_2^4= \frac{V_{12}^2}{V_{22}-V_{11}} (s_1^2-s_2^2)^2
\end{equation}
which can be interpreted as a quadratic equation for $s_1^2/s_2^2$ with
solution
\begin{equation}
 \frac{s_1^2}{s_2^2}= \frac{(V_{22}-V_{11})\sqrt{V_{11}V_{22}-V_{12}^2}-
   V_{12}^2}{V_{11}(V_{22}-V_{11})-V_{12}^2}\,.
\end{equation}
(There is a unique sign choice implied by $s_1^2/s_2^2>0$.)

This solution implies
\begin{eqnarray}
 \frac{s_1^2+s_2^2}{s_1^2-s_2^2}&=&
 \frac{(V_{22}-V_{11})\left(\sqrt{V_{11}V_{22}-V_{12}^2}+V_{11}\right)-
   2V_{12}^2}{(V_{22}-V_{11})\left(\sqrt{V_{11}V_{22}-V_{12}^2}-V_{11}\right)}\\
V_{11}\frac{s_1^2}{s_2^2}-V_{22} &=&
(V_{22}-V_{11})\sqrt{V_{11}V_{22}-V_{12}^2}
\frac{V_{11}-\sqrt{V_{11}V_{22}-V_{12}^2}}{V_{11}(V_{22}-V_{11})-V_{12}^2} 
\end{eqnarray}
which can be used in (\ref{Vbeta2}) to obtain
\begin{equation}
 s_2^4= \frac{\hbar^2}{4}
 \frac{V_{11}V_{22}-V_{12}^2-V_{11}^2}{V_{11}V_{22}-V_{12}^2}
 \frac{V_{11}+\sqrt{V_{11}V_{22}-V_{12}^2}}{(V_{22}-V_{11})\sqrt{V_{11}V_{22}
     -V_{12}^2}+ V_{11}V_{22}-V_{11}^2-2V_{12}^2}\,.
\end{equation}
We also have
\begin{eqnarray}
s_1^4&=&s_2^4(V_{11}\leftrightarrow V_{22})\\
&=&\frac{\hbar^2}{4}
 \frac{V_{11}V_{22}-V_{12}^2-V_{22}^2}{V_{11}V_{22}-V_{12}^2}
 \frac{V_{22}+\sqrt{V_{11}V_{22}-V_{12}^2}}{(V_{11}-V_{22})\sqrt{V_{11}V_{22}
     -V_{12}^2}+ V_{11}V_{22}-V_{22}^2-2V_{12}^2}\,,
\end{eqnarray}
and the angle $\beta$ can be obtained by (\ref{sin2}).

If we insert these solutions in the effective potential, the results can be
seen to equal the low-energy effective potential \cite{EffAcQM}
\begin{eqnarray} \label{Vefffinal}
V_{\mathrm{low}}(q_1,q_2)&=&V(q_1,q_2)+
\frac{\hbar}{2}\sqrt{\frac{1}{2}\left(V_{11}+V_{22}+
\sqrt{\left(V_{11}-V_{22}\right)^2+4
      V_{12}^2}\right)}\\ 
&&+\frac{\hbar}{2}\sqrt{\frac{1}{2}\left(V_{11}+V_{22}
-\sqrt{\left(V_{11}-V_{22}\right)^2+4 V_{12}^2}\right)}\,.
\end{eqnarray}
although it initially appears in a rather different algebraic form. Our
derivation automatically provides results for the ground-state variances and
covariance at the minimum of the effective potential. For instance, while the
actual expression for $\beta$ is quite complicated and not given here, for
small $V_{12}$ we can use a Taylor expansion and obtain
\begin{equation}
\beta=\frac{\pi}{2}+\frac{V_{12}}{(V_{11}V_{22})^{1/4}\left(\sqrt{V_{11}}
+\sqrt{V_{22}}\right)}+O(V_{12}^2)\,. 
\end{equation}
In the limit of weak coupling, the moment $\Delta(q_1 q_2)$ therefore goes to
zero. 

As a simple example, consider the Hamiltonian
\begin{equation}
H=\frac{1}{2}\pi_1^2+\frac{1}{2}\pi_2^2+
\frac{\omega^2}{2}q_1^2+\frac{\omega^2}{2}q_2^2+\gamma 
\omega^2 q_1q_2\,. 
\end{equation}
Its quantization has the exact ground-state energy
\begin{equation}
E=\frac{1}{2}\hbar \omega \left(\sqrt{1+\gamma}+\sqrt{1-\gamma}\right)
\end{equation}
agreeing with what we get from (\ref{Vefffinal}).

\section{Discussion}

Our extensions of canonical variables for moments from second order for a
single degree of freedom demonstrate several new features of semiclassical
states and their dynamics. In particular, we have identified various
parameters related to the impurity of a state, a result which also plays a
role in the determination of semiclassical potentials. Canonical moment
variables are therefore useful tools to understand features of the quantum
state space.

Our other applications illustrate the fact that canonical mappings of the form
derived here can be relevant in a large set of different physical fields. For
instance, they allow one to rewrite quantum statistics in classical terms and
thereby provide convenient access to new types of variables
(Section~\ref{s:thermo}). Interestingly, there is a well-defined partition
function for second-order moments even though these variables are subject to a
non-invertible Poisson structure. For a derivation of the correct phase-space
volume element it is therefore crucial to identify Casimir--Darboux
variables. Casimir variables do not have momenta and therefore do not
contribute the usual $2\pi \hbar$-volume to a partition
function. Nevertheless, in our example we saw that we have to integrate over
them in order to obtain the correct thermodynamical results for fluctuations.

In tunneling situations, canonical moment variables demonstrate a new
heuristic picture of tunneling in which an external field literally opens up a
tunnel through a higher-dimensional extension of the classical potential
(Fig.~\ref{fig:Contourplot}). During tunneling, higher than second-order
moments are crucial, which we have captured by the new all-orders effective
potential (\ref{AllOrders}) defined here for any classical potential. A
separate paper \cite{Ionization} provides a detailed application to tunneling
ionization in atoms with a successful comparison with recent discussions of
experimental results, for which the closure conditions discussed here provide
the foundation.

\section*{Acknowledgements}

This work was supported in part by NSF grant PHY-1607414.


\end{document}